\documentclass{aa}
\usepackage{graphicx}
\usepackage{txfonts}
\bibpunct{(}{)}{;}{a}{}{,} 

\def\kms {km\,s$^{-1}$}

\begin{document}

   \title{Three-dimensional imaging of convective cells in the photosphere of Betelgeuse\thanks{Based on observations obtained at the T\'elescope Bernard Lyot
(TBL) at Observatoire du Pic du Midi, CNRS/INSU and Universit\'e de
Toulouse, France.}}

   \author{{ A.~L{\'o}pez Ariste}\inst{1},{ S.~Georgiev}\inst{2,3}, { Ph. Mathias}\inst{4},  { A. L\`ebre}\inst{2}, { M.~Wavasseur}\inst{1}, { E. Josselin}\inst{2}, { R. Konstantinova-Antova}\inst{3}, { Th.~Roudier}\inst{1}}
   \date{Received ...; accepted ...}

   \institute{IRAP, Universit\'e de Toulouse, CNRS, CNES, UPS.  14, Av. E. Belin. 31400 Toulouse, France \and
   LUPM, Universit\'e de Montpellier, CNRS, Place Eug\`ene Bataillon, 34095 Montpellier, France \and
   Institute of Astronomy and NAO, Bulgarian Academy of Science, 1784 Sofia, Bulgaria \and
   IRAP, Universit\'e de Toulouse, CNRS, UPS, CNES, 57 avenue d'Azereix, 65000, Tarbes, France
}

 
  \abstract
  {}
   {Understanding convection in red supergiants and  the mechanisms that trigger the mass loss from these evolved stars are the general goals of most observations of Betelgeuse and its inner circumstellar environment.}
   {Linear spectropolarimetry of the atomic lines of the spectrum of Betelgeuse reveals information about the three-dimensional (3D) distribution of brightness in its atmosphere. We model the distribution of plasma and its velocities and use inversion algorithms to fit the observed linear polarization.}
   {We obtain the first 3D images of the photosphere of Betelgeuse. Within the limits of the used  approximations, we recover vertical convective flows and measure the velocity of the rising plasma at different heights in the photosphere. In several cases, we find this velocity to be constant with height, indicating the presence of forces  other than gravity acting on the plasma and counteracting it. In some cases, these forces are sufficient to maintain plasma rising at 60\,\kms to heights where this velocity is comparable to the escape velocity.}
   {Forces are present in the photosphere of Betelgeuse that allow plasma to reach velocities close to the escape velocity. These mechanisms may suffice to trigger mass loss and sustain the observed large stellar winds of these evolved stars. }
  

\titlerunning{Convective cells in Betelgeuse}
\authorrunning{A. L\'opez Ariste et al.}

   \maketitle

\section{Introduction}
 Betelgeuse is a very interesting target for the observation and understanding of red supergiants (RSG). The study of the convective movements in these cold, low-gravity stars advances our understanding of turbulent fluid motions \citep{freytag_spots_2002,chiavassa_radiative_2011}. This very same turbulence may give rise to the magnetic fields observed in Betelgeuse \citep{auriere_magnetic_2010,Mathias:2018aa}. Furthermore these convective motions in the photosphere may largely contribute to the increased mass loss at these stages of the stellar evolution, forming a strong stellar wind which  greatly contributes to the chemical enrichment of the circumstellar and interstellar medium.

Its large angular diameter  \citep[42.11 mas in the K band,][] {montarges_dimming_2021}, due to its relative proximity (about 200\,pc) but also to its actual size as an MIab supergiant, has made  Betelgeuse a favorite target for interferometry, as these characteristics allow  detailed images of its photosphere which have unveiled large convective structures \citep[e.g.,][]{Haubois2009,montarges_close_2016}. These structures were recently confirmed through an unexpected technique based on the discovery \citep{Auriere_2016} and interpretation \citep{LA18} of linear polarization in the atomic lines of the spectrum of Betelgeuse. 
The analysis of these signals allowed \cite{LA18} to produce images of the star.  These images were comparable to quasi-simultaneous interferometric images and unveiled large convective structures that we can call granulation, as in the Sun. Another result of this new technique was the observation of large convective velocities, of about 40\,\kms, well above the predictions in the adiabatic approximation, but in agreement with the results from  numerical simulations by
\cite{freytag_spots_2002} and \cite{chiavassa_radiative_2011}.

The amount of linear polarization observed in these atomic lines is often of few hundredths of a percent, and is therefore below the typical signal-to-noise ratios (S/Ns) of single observations. It has been customary in stellar spectropolarimetry to add the signals  of a large number of atomic lines in order to increase the S/N of the resulting mean spectral line \citep{donati_spectropolarimetric_1997}. Adding up different spectral lines from different atomic species may sound incongruous,  but it is acceptable if the polarization signal is expected to be the same, up to a scale factor, for all the lines added. \cite{Auriere_2016} showed that this was indeed the case for the linear polarization of Betelgeuse. The images inferred by \cite{LA18} depend upon the interpretation of the linear polarization of this mean line, which is the result of the addition of over 10 000 atomic lines. However one can do better than adding up all those lines regardless of the properties of each individual line.  \cite{Auriere_2016} already produced a figure (their Fig.\,4) where the line addition was made over lines with similar depth in the intensity line profile. 
Six masks were thus produced which included lines with a line depth (the depression of the absorption line with respect to the continuum)  of between 0.9 and 1, or those between 0.8 and 0.9, and so on, depths which were computed in a Vienna Atomic Line Data Base (VALD) model atmosphere \citep{Auriere_2016}.  
The authors did not advance an interpretation of the profiles shown in their figure, but it is  apparent that those profiles, while similar, are not identical. 
Furthermore, the differences cannot be attributed to noise. 
First, the amplitude changed in such a way that the deeper the line, the larger the amplitude of linear polarization. 
A more subtle result was that the maximum of emitted polarization did not appear at the same wavelength for different line depths.  It is clear that these lines, grouped by the depth of their intensity line profile, are not a signature of the same photospheric structures. It is also obvious that, as line depth can be roughly seen as a proxy of height of formation, these lines contain information from different heights in the atmosphere.     The present work stems from those observations and infers the first three-dimensional (3D)  images of the photosphere of Betelgeuse.
Tomography based on contribution functions is a well-known technique \citep{kravchenko_tomography_2018}, even for the case of Betelgeuse and other red supergiants. In this work, we are not doing it on the intensity but instead on the polarization spectra. As discussed below, polarization shows particularities that render most of the previous work on tomography essentially irrelevant in the present case. These particularities force us to re-derive the conditions and constraints that apply to our data.

In Section 2 we present our large set of observations collected from 2013 to 2021 with the spectropolarimeters Narval and its upgraded version Neo-Narval at the Telescope Bernard Lyot at Pic du Midi (France).
In Section 3, we discuss the many approximations needed to produce those 3D images. 
Those approximations prevented a straightforward interpretation of the inferred images in the two-dimensional (2D) case, and indeed they impose even stronger limits on the interpretation of 3D images. 
Nevertheless, in Section 4 we present two cases from data from 2013 and 2020 observed with Narval and its upgraded version Neo-Narval, respectively. The inferred images reveal interesting aspects of the convective movements in the photosphere of Betelgeuse in spite of all the approximations made. 
In Section 5, we focus on the dependence of the velocity of the rising plasma on height. 
We isolate particular cases in which the plasma rises at constant velocity. 
The large velocities measured, 40 and even 60\,\kms\, in some cases, are kept constant up to heights where they are comparable to, but smaller than, the escape velocity.

\section{Spectropolarimetric data from  Narval and Neo-Narval}

We have been monitoring Betelgeuse in linear polarimetry since November 2013 with the Telescope Bernard Lyot at Pic du Midi (France,TBL). Until August 2019, the Narval spectropolarimeter was used, and this data set has been described by \cite{Auriere_2016,Mathias:2018aa} and \cite{LA18}. In September 2019, Narval was significantly upgraded and became Neo-Narval, and the first data for Betelgeuse were collected by February 2020.

Neo-Narval maintains the successful polarimeter of Narval \citep{Donati2006}, but has been upgraded with a new, higher performance detector, the iXon CCD by ANDOR. 
The faster readout times of the new detector (2 to 3 seconds) at  readout noises comparable to the old detector of Narval (2 to 3 electrons in the usual configuration) and similar quantum efficiencies allow a more efficient use of telescope time. 
The spectrograph maintains the main performances of Narval, that is, a spectral coverage from 380nm through 1050nm, and a median spectral resolution of 65 000 after data reduction, but  has been thermally stabilised with three concentric enclosures of which the middle one is actively controlled in temperature. 
The diffraction grating has been mounted inside an isobaric chamber. 
Altogether, the goal is a stabilisation of the spectrograph to allow it to measure velocities with a precision of 3\,m\,s$^{-1}$, but  thermal and pressure control of the spectrograph are not sufficient.  The final sensitivity to velocities is afforded by the introduction of a calibration beam fed with a stable Fabry-Perot. 
Each one of the 40 spectral orders seen by Neo-Narval is thus split in three: two beams carry the orthogonal polarizations that allow the polarization modulation, while the third one carries the Fabry-Perot spectrum. 
In order to introduce this third calibration beam, the camera optics and the cross-dispersing prisms were overhauled. 
As a result, the raw image of the spectral orders differs considerably from the previous one with Narval, and required an upgrade of the data-reduction software.  
Rather than refurbishing the old Libre-Esprit code \cite{donati_spectropolarimetric_1997}, Neo-Narval engaged in providing a completely new code written from scratch. 
For the present work, the modifications in Neo-Narval concerning velocity measurements are of no importance. 
However, the new DRS  had to be validated for our present purposes.
Since January 2020, three major upgrades of the DRS have been implemented, and the data re-calibrated with the latest version of the code. 
Although the DRS still shows room for improvement in terms of velocimetry and the normalization of the intensity continuum signal, the  tests with respect to Betelgeuse spectropolarimetry, which are  our only concern here, show a continuity with Narval data. The amplitude of the polarization signals, the S/Ns  reached in polarization and instrumental broadening, all appear to rest unchanged from one instrument to the other. 
Some of the plots in this paper (Figs. \ref{QUmasks}, \ref{lms13}, \ref{lms21} and \ref{cuts}) presenting measurements made directly on the profiles are particular examples of this continuous quality of the data.

For completeness, a list of observations of Betelgeuse made with Narval (August 2018 - August 2019) and Neo-Narval (February 2020 - February 2021) is presented in Table\,1. We note that the upgrade of Narval to Neo-Narval resulted in an interruption of several months of our regular monitoring of Betelgeuse that was also very unfortunately coincident with its famous episode of great dimming in the period December 2019-February 2020 \citep{montarges_dimming_2021}.

\section{Approximations and assumptions}

The 2D images produced by \cite{LA18} are the result of fitting the observed linear polarization profiles with synthetic ones obtained from a model. 
The basic idea behind these reconstructed images is that  the observed polarization is due to the nonuniform distribution of brightness over the stellar disk. 
The spatial distribution of the brightness can be inferred by considering that, due to Rayleigh scattering, the polarization emitted by a bright spot over the disk will be tangent to the local limb, and therefore its ratio of Stokes Q to Stokes U linear polarization signals will provide the polar angle over the disk of such a spot. 
The distance of that spot to the center of the disk can be inferred from the wavelength at which a linear polarization signal is found.  
Assuming no rotation, the presence of convective motions will Doppler-shift the signals in such a way that a bright spot found near disk center will send its polarization to the blue wing of the atomic line, while a spot near the limb will send its signal onto the red wing, irrespective of its polar angle.
Although analogous, we note that this projection of disk positions onto wavelength is completely different from the more familiar projection of a rotationally broadened line.
This basic idea can be extended to more complex scenarios with the help of an inversion algorithm and of a model for the continuous distribution of brightness and velocities over the disk, as described by \cite{LA18}. 
This method to infer images nevertheless rests upon a generous number of approximations and assumptions about the structure of the photosphere of Betelgeuse, radiative transfer, and the emission of polarization.  
All of these approximations and assumptions were vindicated by the observed similarities between the resulting images and those resulting from interferometric observations.   
Furthermore, it appeared that the  main conclusions of that work were not fundamentally altered by them: the spatial scale of the observed structures was constrained by the spectral width of the observed polarization signals; the temporal scales were independently corroborated through time-frequency analysis of the spectra by \cite{Mathias:2018aa}; and the maximum convective velocities found were forced by the wavelength span of the observed polarization signal.

This state of affairs cannot be safely pursued when  also trying  to extract height information from the data. 
An examination of the validity and impact of those approximations is required before we trust the 3D images produced in the present work. 
The basic approximation taken is that polarization is emitted by the last scattering in a gray atmosphere. 
Under this assumption, we can order the atomic lines in terms of height of formation as we show below, but we are unable to compute its height. 
Next, we should assume certain values for both the heliocentric radial velocity of Betelgeuse $V_*$ and the maximum upward speed of the convective fluxes $V_{max}$. 
Both parameters are constrained by observations, but a definite value cannot be measured, but simply estimated. While there is a huge amount of information in the polarized profiles, it is not sufficient to unambiguously image the photosphere of Betelgeuse, and therefore there is a random choice of one solution. 
We  examine these approximations one by one.

\subsection{Maximum and minimum velocities}


Betelgeuse is modeled as a convective, nonrotating star. 
This implies that bright, hot plasma is rising, emitting blueshifted profiles, while dark, cold plasma is sinking, emitting redshifted profiles. By making this explicit description of the contribution of the convection velocity fields to the spectral line profile we exclude these velocity fields from contributing to  macroturbulence. Macroturbulence will still contain any turbulent fields associated to convection, as well as any other velocity contribution at scales larger than the mean free photon path.
Therefore, in the absence of rotation, these convective velocities are the only macroscopic velocity fields present. 
The net addition of spectral line profiles over the stellar disk (ignoring 3D radiative transfer effects) will be dominated by the bright regions. 
From this argument, we should expect a net intensity line profile which is blueshifted with respect to the radial velocity of the center of mass of the star relative to us.   For a convective non-rotating star,
the  center of the intensity line profile is not  a measure of this radial velocity. 
If only convective velocities were present, this radial velocity, $V_*$, of the center of mass  of the star in the heliocentric reference system would lie somewhere in the red wing of the observed line. 
Where it lies exactly depends on the contrast ratio between the bright rising  plasma and the dark sinking plasma, and of the inhomogeneities of brightness at different parts over the visible disk. 
This radial velocity is a parameter that should be fixed before  the inversions are done.

If our model had included the presence of rotation, it would  introduce a symmetrizing effect on the atomic line.  Were rotation velocities to dominate the profile shape, the true velocity $V_*$ would be somewhere near the line center. A similar reasoning can be made for other line-broadening mechanisms: the eventual presence of macroturbulent velocity distributions  would tend to symmetrize the profile around the true zero velocity of the center of mass.  Determining this velocity $V_*$ of the center of mass of the star  therefore requires the contribution of these 3 very different velocities to the line profile  to be disentangled: convective velocities that tend to place $V_*$ in the red wing of the observed intensity line profile, and rotational and thermal velocities that tend to shift it toward the center of that profile. The presence of narrow lobes in the profiles of linear polarization, as illustrated in the profiles of Fig.\ref{QUmasks}, is proof that thermal, micro- and macro-turbulent broadening are much smaller than the span of values attributed to convective velocities. The estimated rotational periods of Betelgeuse \citep[in the range of 12 to 30 years,][]{uitenbroek_spatially_1998,kervella_close_2018} result in maximum rotation velocities of at most 5 or 10\,\kms. Such rotational velocities are  comparable to the thermal broadening but, again, much smaller than the observed span of velocities (40 - 60 \kms) seen in the polarization signals. For these {\it a priori} reasons, our model neglects rotation and introduces an {\it ad hoc} broadening that also takes into account the instrumental spectral resolution. This broadening is included as a convolution of the signal by a Gaussian profile with FWHM  10\,\kms; this is an empirical value that nicely corresponds to the estimated macroturbulence from 3D simulations \citep{chiavassa_radiative_2011}.
After all these considerations, we expect $V_*$ to lie somewhere on the red side of the observed intensity line profile, but its true position is unknown and therefore constitutes a parameter of our model.

The previous discussion stems from the comparison of our model of a convective nonrotating star with what is observed in the spectral profiles of linear polarization of atomic lines in the spectrum of Betelgeuse. \cite{Auriere_2016} and \cite{LA18} based their results and produced images comparable to concurring interferometric observations, on such a model constrained by linear polarization alone.  The line profiles in intensity were left aside because, as we examine in detail below,  intensity line profiles are in clear disagreement with observed polarization profiles.
Polarization profiles show large amplitudes in the far wings of the intensity line profiles (an illustration of this can be found in the data  for the dates retained for the present work which will be introduced and described later; Fig.\,\ref {QUmasks}). 
As a result, the polarization appears to emerge from an atomic line many times broader than the observed intensity line profile. 
Our current framework for the interpretation of those polarization signals cannot justify this fact. 
A simple addition of the intensity  over the 2D images produced by \cite{LA18} results in intensity line profiles that are much broader than the observed ones. Adding the effects of limb darkening, for example, to this simple addition of profiles  does not help, because our model sends the signal from disk center to the blue end of the line. Including limb darkening will simply shift the computed intensity profile to the wing of the observed profile.
It could be argued that the present model used to fit the linear polarization profiles is simply incorrect because it does not  also fit the intensity line profiles. We argue below that the ability of the present model to produce images comparable to those from interferometric observations of Betelgeuse and also of CE Tau, is a strong backing for the model fitting linear polarization. This model also inferred the right spatial scales and timescales of the convective structures, and measured the high velocities of the plasma \citep{LA18} that are predicted by 3D simulations. None of this has been achieved by looking exclusively into the intensity line profiles.  In contrast, no spatial scale has ever been derived from the analysis of intensity profiles, and even a thorough analysis of their variability produces velocities   which are limited to simply a few km/s \citep{Gray2008} in disagreement with theoretical models and our measurements based on polarimetry. This state of affairs lends credit to the model able to interpret linear polarization profiles and forces us to try to understand why the profile of the atomic lines in intensity disagrees with the observed linear polarization.
One possibility emerges  when  one takes into consideration the fact that most of the radiative transfer takes place in a moving atmosphere \citep{chandrasekhar_formation_1945,bertout_line_1987}. The recent example of Mira stars illustrates that radiative transfer through the moving shells - ejected during present and previous pulsation cycles - is subject to a geometry weighing that favors the center of the stellar disk. The resulting disk-integrated profile is not the simple addition of the local intensity line profiles over the whole disk, but just over a much smaller region around the disk center \citep{bertout_line_1987,lopez_ariste_asymmetric_2019}. Were a similar phenomenon  at work in Betelgeuse, the observed intensity line profile would correspond to a reduced region of the disk, justifying its narrow width and its lack of consistency with the polarization profiles.  \cite{lopez_ariste_asymmetric_2019} demonstrated that, in the case of the expanding atmosphere of the Mira star $\chi$ Cyg, the linear polarization was also affected by a geometric factor composed of the product of the same geometric weight in the integral of the intensity over the stellar disk, which favored the disk center, but also  of a further $\sin^2 \theta$ weight typical of scattering polarization. Polarization was, unlike intensity, sensitive to the whole disk.

\begin{figure}[htbp]
\includegraphics[width=0.5\textwidth]{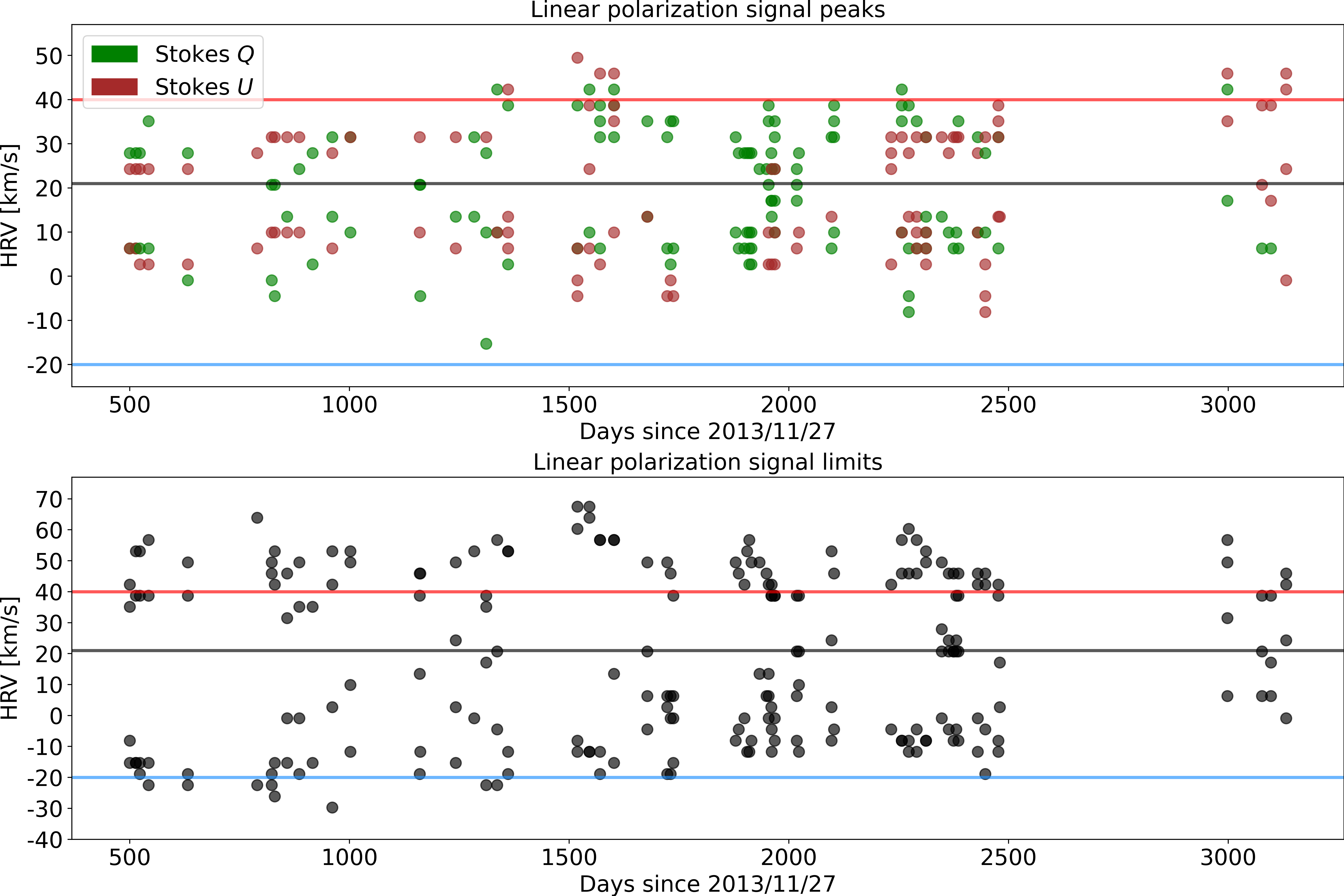}
	 \caption{Determination of the position of the observed peaks in polarization (top) and the maximum wavelength at which signal is still above noise (bottom) for all the data available since 2013. The gray horizontal line marks the average wavelength of the intensity line center.  The blue and red lines mark the chosen values of $V_{max}=60$ \kms\, and $V_*= 40$ \kms\, respectively. Wavelengths are given in \kms\, in the barycentric reference.  Neo-Narval data start from day  2700. Noise is defined and measured as the standard deviation of the polarization signal at wavelengths beyond 100\kms of the center of the spectral line. Typical noise values are below $10^{-4}$ of the continuum intensity.}
\label{velsProf}
\end{figure}

At present, it is not clear whether or not radiative transfer through a moving atmosphere is able to explain the differences seen between the intensity and polarization profiles in Betelgeuse. 
In what follows, we  base our interpretations on the information provided by polarization line profiles alone, and assume that the intensity line profile has somehow been restricted to represent a small portion of  the disk, and does not provide complete information on the convective velocities present in the atmosphere of Betelgeuse. 
To determine $V_*$, we must rely  only on the polarization signals. 
Figure\,\ref{velsProf} shows the velocities in the heliocentric reference frame of both the peaks and the wings of the  polarization signals observed since 2013 and extracted from line addition of the full set (about 10 000) of atomic lines available.
The horizontal gray line in both plots shows the center of the intensity line.  
The ordinates refer to the barycentric frame. 
In this reference system, the estimated radial velocity $V_*$ of the Betelgeuse center of mass is redshifted. 
Its value of $V_*=40\,$\kms\, is  chosen so that most of the polarization peaks   are blueshifted, as expected from spectral profiles dominated by convective flows. 
The reason for not pushing $V_*$ to a more extreme value is that we leave a margin of about $10 \,$\kms,  allowing for some redshifted profiles.
While very few peaks appear redshifted with this choice, the bottom plot in Fig. \ref{velsProf} shows that there is a non-negligible amount of redshifted signal. 
We attribute these redshifted signatures  mostly to the dark, sinking plasma. Conversely, choosing a smaller value of $V_*$, and closer to the center of the intensity line profile  would leave a large part of the observed linear polarization signal unexplained. Within the cited margin of $10 \,$\kms, the velocity of the center of mass cannot be significantly different than the one chosen if our model is to fit the observations.
With this choice of $V_*$, the observed convective velocities span a range of at least 40 to 50$\,$\kms, with some examples at even 60$\,$\kms.  As we mention above this large range allows us to discard rotation as a primary source of line broadening.  
Similar arguments can be made for thermal, micro-, and macroturbulent velocity fields.

It may be tempting, in spite of the {\it a priori} arguments given above and  in view of these two plots in Fig.\,\ref{velsProf}, to return to a symmetric distribution of velocities with $V_*$ at the center of the intensity line profile: the linear polarization signals appear symmetrically distributed with respect to the observed center of the intensity line profile. 
To counter this possibility, we produce a last argument: we must recall that \cite{LA18} inferred an image of CE\,Tau with a bright granule at disk center, concurrent with simultaneous interferometric images of \cite{montarges_convective_2018}. 
In the current framework, the presence of such a bright granule at disk center was inferred from  the observation of a strong linear polarization signal at the most blueshifted wavelengths. 
The reality of such a structure forces us to conclude that, first, there is spatial information in the wavelength distribution of polarized signals and, second, that disk center emits signals at the bluest wavelengths and not at the center of the  line profile. 
Any other modelling that would deny a relationship between wavelength and spatial position (as macroturbulent velocities do) or that would call for disk center to emit at the line center wavelength (as rotational velocities do) must be excluded.

Having fixed the value of $V_*$ in the red wing of the line at 40$\,$\kms\,  within a margin of error of about 10\,\kms, we turn our attention to the span of velocities present in the linear polarization signals. 
Our model assumes that the rising plasma can have a maximum velocity $V_{max}$. 
The determination of this second velocity parameter $V_{max}$ is of utmost importance for our model and our conclusions.  
Due to the relationship between velocity and wavelength in the present model, with just convection and without rotation, a bright granule strictly at disk center emits its signal at the bluest wavelength, corresponding in velocity coordinates to $V_* - V_{max}$.
Inspection of Fig.\,\ref{velsProf} shows that, over the last 7 years, peaks in polarization have been observed up to 60\,\kms\, with respect to the star's  radial  velocity $V_*$.  
For  signals in the wings that are above the noise  we should have considered even larger maximum convective velocities $V_{max}$:  beyond 60\,\kms\, in the stellar reference frame. 
Altogether, leaving a margin of 10\,\kms , we can establish a maximum velocity for the rising plasma of $V_{max}=60$\,\kms. 
Such high velocities of rising plasma seem to be  justified by numerical simulations \citep{chiavassa_radiative_2011}.

Even if we can, from data, measure the maximum observed velocity $V_{max}$ of the convective flows, we cannot expect this maximum velocity to be present at a given moment on Betelgeuse. 
However, we cannot afford to have a different value of $V_{max}$ for every observing date: this would add an extra parameter to the model and we cannot ensure that there is a bright granule at disk center on every date to send a signal at the bluest wavelength that fixes such a parameter. As adding a free parameter for which information is probably not always available is impractical, we fix a constant $V_{max}$ for our whole dataset.
The relationship that our model establishes between wavelength and position over the disk implies that any bright plasma at disk center is assumed to move at velocity $V_{max}$. 
If at a particular date the plasma velocities present over the star are lower than this maximum $V_{max}$, the inversion algorithm will not place any bright structure at the center of the disk. 
The inferred bright structures will be placed off center just because their velocity is smaller than the maximum velocity. 
This is an important bias to keep in mind when analyzing the inferred images of Betelgeuse, and a bias that shall be exploited later on. 
In the absence of any independent method to fix the maximum velocity for every individual date, this is a difficult limitation to resolve.

\subsection{Gray  atmosphere}

To simplify the computation of the emergent polarization $Q(\nu)$ in the direction of the observer,  we   assume that all polarized photons are due to scattering, and not to any process of dichroism. This allows us to write, following \cite{landi_deglinnocenti_polarization_2004}, that
\begin{equation}
Q(\nu)= \frac{3}{2\sqrt{2}}\left(1-\mu^2\right)\int^{\infty}_0 \beta J_0^2(\nu) e^{-\frac{t_\nu}{\mu}}\frac{dt_\nu}{\mu}
\label{grayeq}
\end{equation}
with $\beta$ the fraction of opacity due to scattering, $J_0^2(\nu)$ the irreducible spherical tensor   of the illumination as a function of $\nu$, the frequency of light, $\mu$ the distance to disk center, and $t_\nu$ an integration variable that can be seen as a frequency-dependent optical depth.
It is useful to introduce the anisotropy factor
\begin{equation}
w_{\nu}=\sqrt{2}\frac{J_0^2(\nu)}{J_0^0(\nu)}
\end{equation}
and rewrite the previous expression as 
\begin{equation}
Q(\nu)= \frac{3}{4}\left(1-\mu^2\right) w_{\nu} \int^{\infty}_0 \beta J_0^0(\nu) e^{-\frac{t_\nu}{\mu}}\frac{dt_\nu}{\mu}=\frac{3}{4}\left(1-\mu^2\right) w_{\nu} I_{scatt}(\nu)
\label{grayeq2}
\end{equation}
At any given position over the disk of  Betelgeuse, the emergent polarization will vary mostly following  this anisotropy factor $w_{\nu}$. In particular, at $\mu=0$ we have
\begin{equation}
\frac{Q(\nu)}{I_{scatt}(\nu)} \approx w_{\nu}
\label{fract}
\end{equation}

Following  \cite{landi_deglinnocenti_polarization_2004} closely, where details of these calculations can be found, we can compute the anisotropy factor $w$ in two extreme cases: scattering in the outer atmosphere and in a gray atmosphere. 

In the case of scattering in the outer atmosphere, the scattering event takes place sufficiently far away from the star that the illumination is cylindrically symmetric around the radial direction with only limb darkening to take into account. 
For distances that are small compared to the radius of the star,  an approximation that appears as suitable for photospheric lines in Betelgeuse, the anisotropy factor $w_{\nu}$ can be written as a correction to the factor at zero height, $w_{\nu,0}$, in terms of the height $h$ and the limb-darkening functions $u_1(\nu)$ and $u_2(\nu)$:
\begin{equation}
w_{\nu}=w_{\nu,0}+\frac{9}{5}\frac{[1-u_1(\nu)-u_2(\nu)][20-5 u_1(\nu)-8 u_2(\nu)]}{[6-3u_1(\nu)-4u_2(\nu)]^2}\sqrt{\frac{h}{2R_*}}
\end{equation}

In the case of scattering in a gray atmosphere, the scattering event takes place in an environment where the emission is thermal and described by a Planck function. The radiative transfer equation for the integrated flux, an approximation to the mean line arising from least-squares deconvolution (LSD), can be solved in terms of its moments, and one can derive an anisotropy factor $w$ for any optical depth $\tau$ in the form
\begin{equation}
w(\tau)=\frac{q(\infty)-q(\tau)}{2[\tau+q(\tau)]}
\end{equation}
in terms of the Hopf function $q(\tau)$.

\begin{figure}[htbp]
\includegraphics[width=0.5\textwidth]{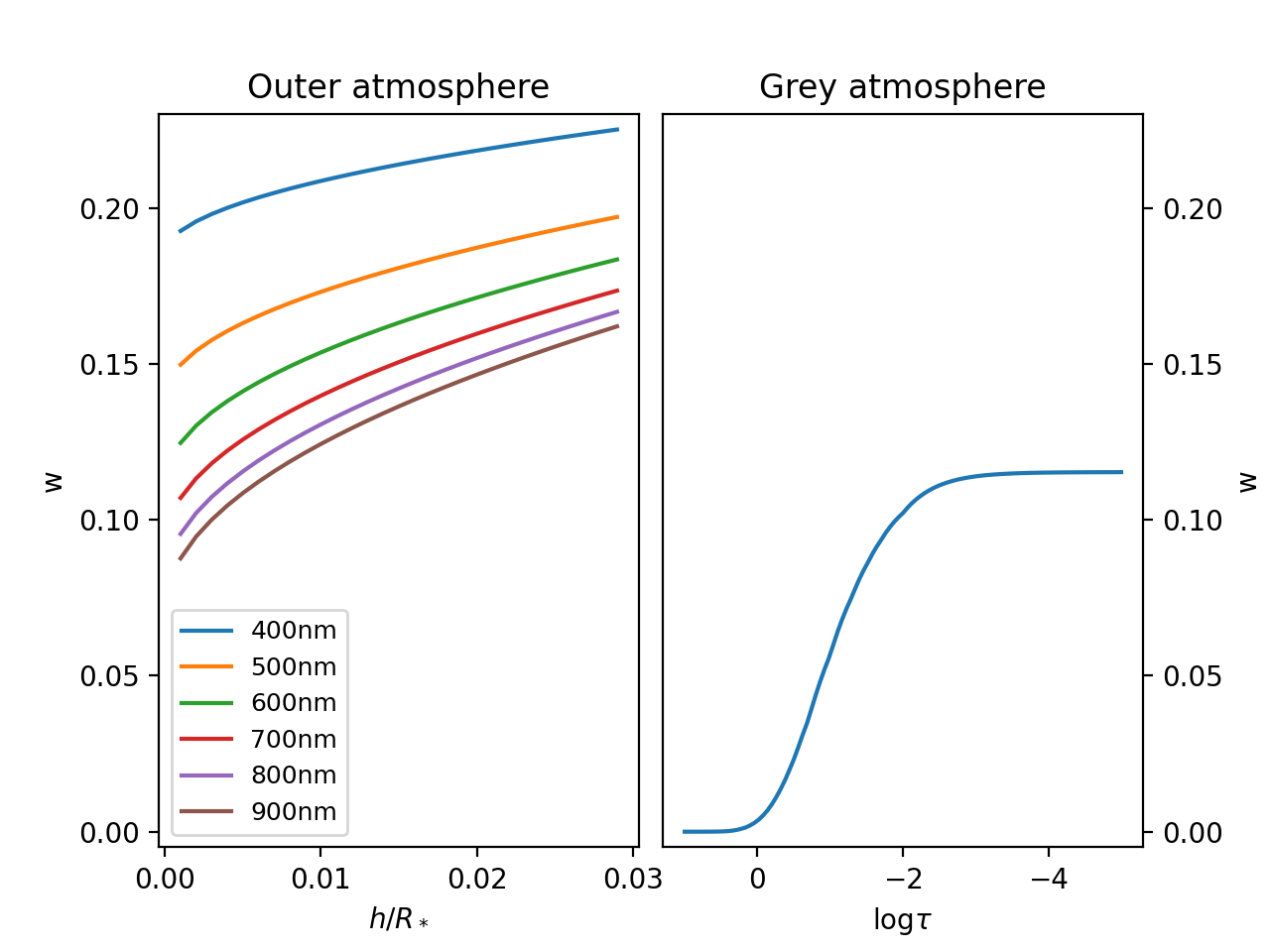}
\caption{Computation of the anisotropy factor for the case of the outer atmosphere (left) and the gray atmosphere (right) approximations.  Because of the nature of those approximations, the abscissae are different: optical depth in the case of the gray atmosphere, and distance to the photosphere in the outer atmosphere case. In this last approximation, computations are shown for six different wavelengths in the spectral domain of Narval.}
\label{w}
\end{figure}

In Fig.\,\ref{w} we compute both cases: in terms of height above the star and a solar-like limb darkening for the extreme outer-atmosphere case, and in terms of optical depth for the case of the gray atmosphere. 
The importance of the figure lies in the extreme difference in the resulting values of the anisotropy factor. 
If one misreads Eq.(\ref{fract}) as an approximation of the observed linear polarization rate, the outer atmosphere scenario  predicts a local polarization amplitude of the order of a few percent  even at very low heights. 
In the gray atmosphere, polarization amplitudes diminish monotonically  to much lower polarization amplitudes as one goes deeper into the atmosphere and radiation arrives isotropically from all directions. We must stress that the only aspect of the gray atmosphere approximation we are using here is the fact that light may come from all directions to the last scattering point, in contrast with the outer atmosphere case, where light comes from a unique direction. This difference alone allows us to reach our conclusions. One could  drop all other approximations linked to the gray atmosphere and reach the same conclusions, but without the benefit of a simple analytical formulation, and bringing in no further insight.

Rather than a demonstration, the previous arguments are simply an educated guess as to what the best description for the emergent polarization of Betelgeuse may be. 
If this polarization is of the order of 10\%, we shall not be able to tell apart the two approximations  without computing them in much more detail. However, if the emitted polarization happens to be of the order of 1\%, as is shown to be the case below,  we will be able to conclude that the gray atmosphere is the right framework.   \cite{Auriere_2016} concluded that the observed polarization is just the polarization of the continuum depolarized by the  atomic lines forming directly above it. Continuum polarization in Betelgeuse has been rigorously computed by \cite{josselin_atmospheres_2015} or \cite{LA18} and seen to be of the order of 1\% at the limb, overseeing wavelength variations, a figure that is in agreement with the few measurements available \citep{1986ApJ...307..261D,doherty_polarization_1986,schwarz_polarization_1984}. If the identification of \cite{Auriere_2016} is correct, then a gray atmosphere is the right scenario, in terms of the anisotropy of the incoming radiation. Further confirmation must come from observations. To compare with observed values we should recall that it is not the local polarization rate that one measures, either in the continuum or in the atomic lines of the observed spectrum, but the disk-integrated polarization. A perfectly homogeneous disk would perfectly cancel out the integrated polarization. The presence of inhomogeneities ensures that there is a nonzero integrated polarization, but  with a reduced amplitude. We can estimate the reduction in polarization amplitude due to this disk integration by looking into the images inferred by \cite{LA18}. In that work, it was concluded that the observed polarization profiles could all be reproduced by  assuming photospheric brightness distributions given in terms of spherical harmonics of a maximum degree of 5. We  launched random combinations of spherical harmonics of such degree to simulate the aspect of Betelgeuse and computed the net polarization expected compared to the initial local polarization.

\begin{figure}[htbp]
\includegraphics[width=0.5\textwidth]{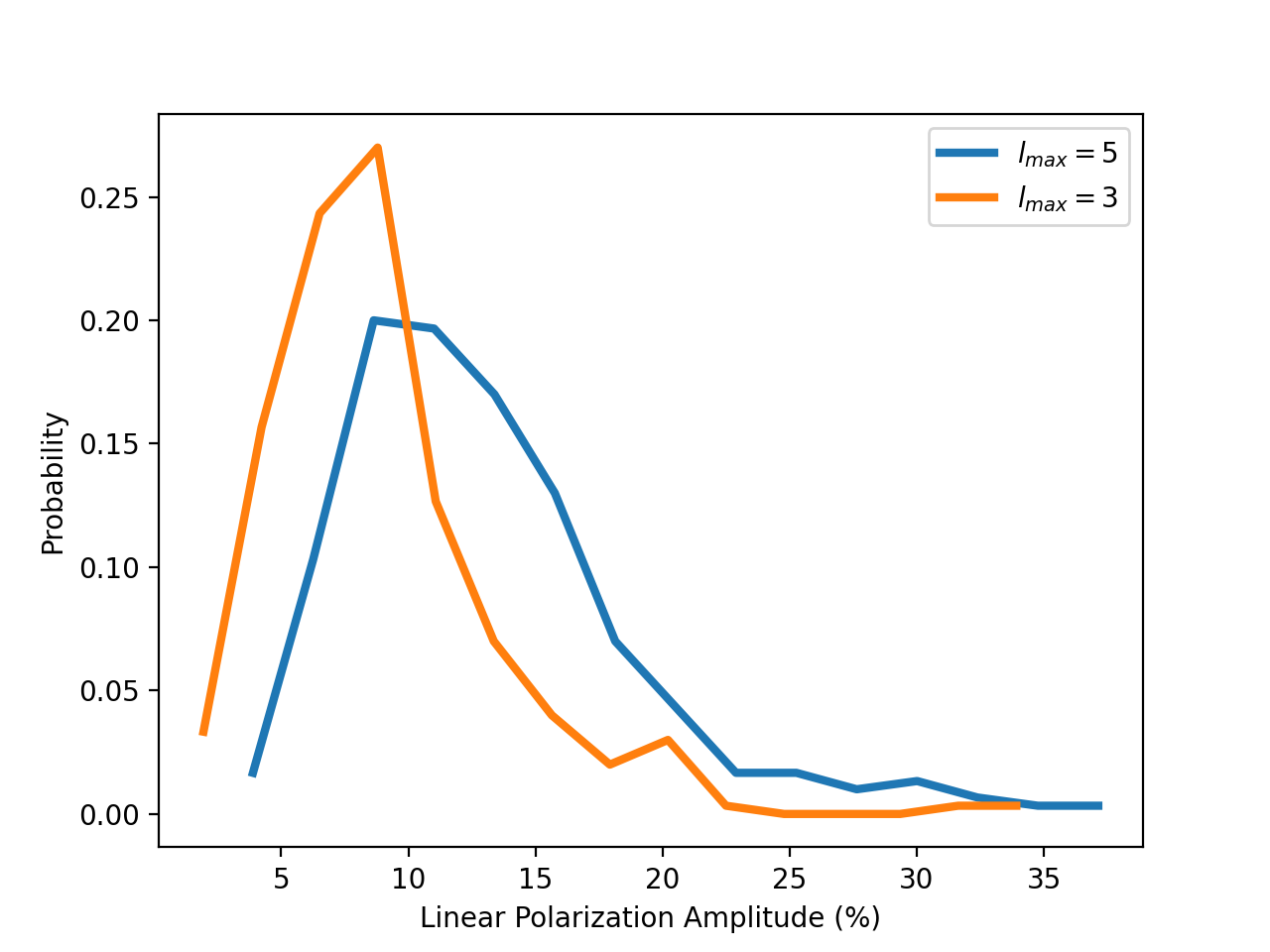}
\caption{Expected polarization rate from different brightness distributions assuming that local scattering results in 100\% polarization. The histograms are built from realistic brightness distributions built with a combination of spherical harmonics of maximum order $l=3$ and $l=5$.   }
\label{gauge}
\end{figure}

Figure\,\ref{gauge} shows histograms of this expected net polarization (the case of maximum degree 3 for the spherical harmonics is also shown). These histograms peak at roughly 10\%. That is, integrating over the disk in the presence of brightness structures of the size of those observed in Betelgeuse roughly reduces the local polarization by a factor of ten to a mere 10\% of the initial local polarization.  We can now 
  go to the observations and see what is the actual net polarization observed. Over 7 years of observations, the maximum amplitudes of polarization rates  in Stokes Q and U profiles are seen to be of 0.05\% to 0.1\%.
  We must conclude that the local polarization in Betelgeuse must have been roughly of 0.5 to 1\% , in close agreement with the radiative transfer computations of the polarization  of the continuum.

A consistent picture appears. Assuming that the origin of polarization is Rayleigh scattering of the continuum, depolarized by atomic lines, we expect a maximum  local polarization of about 1\%. With that same hypothesis, we can model the brightness distribution of the photosphere of Betelgeuse and  fit the observed polarization profiles. The models that fit the observed profiles present bright structures of the size predicted by convective theory in red supergiants \citep{LA18}. Such brightness distributions would prevent linear polarization from canceling out in the disk-integrated spectra, but they would nevertheless diminish the observed net polarization by a factor of ten. Starting from a maximum of 1\%,
we expect that observed polarization should  be of 0.1\%, which is the rough order of magnitude of the observed polarization.  This match of predicted and observed amplitudes is a positive point for the inversion and imaging technique, given that at no point do the inversion algorithms use the result of the expected local polarization amplitudes computed from radiative transfer.

Re-examination of Fig.\,\ref{w} at this point leads us to the conclusion that only the gray atmosphere can be accepted as approximation to the anisotropy of the incoming radiation. 
Indeed, equating the anisotropy factor with emergent polarization, a gray atmosphere would produce local polarization amplitudes of 1\% for $\log \tau$ in the range $(-1,0)$, where the optical depth $\tau$ reaches 1 when light comes from all directions and the anisotropy factor $w\approx 0$.
The outer atmosphere approximation, on the other hand, would require heights of 0.001\,R$_*$ to approach such low polarization amplitudes. 
At these heights,  for Betelgeuse, it is not realistic to impose the approximation of light coming from  just  one direction, as required for  an outer atmosphere, and so  it appears that we must keep the gray atmosphere as the appropriate approximation to compute our expected polarizations.

\subsection{ The single scattering approximation: information along the optical path}
\begin{figure}[htbp]
\includegraphics[width=0.5\textwidth]{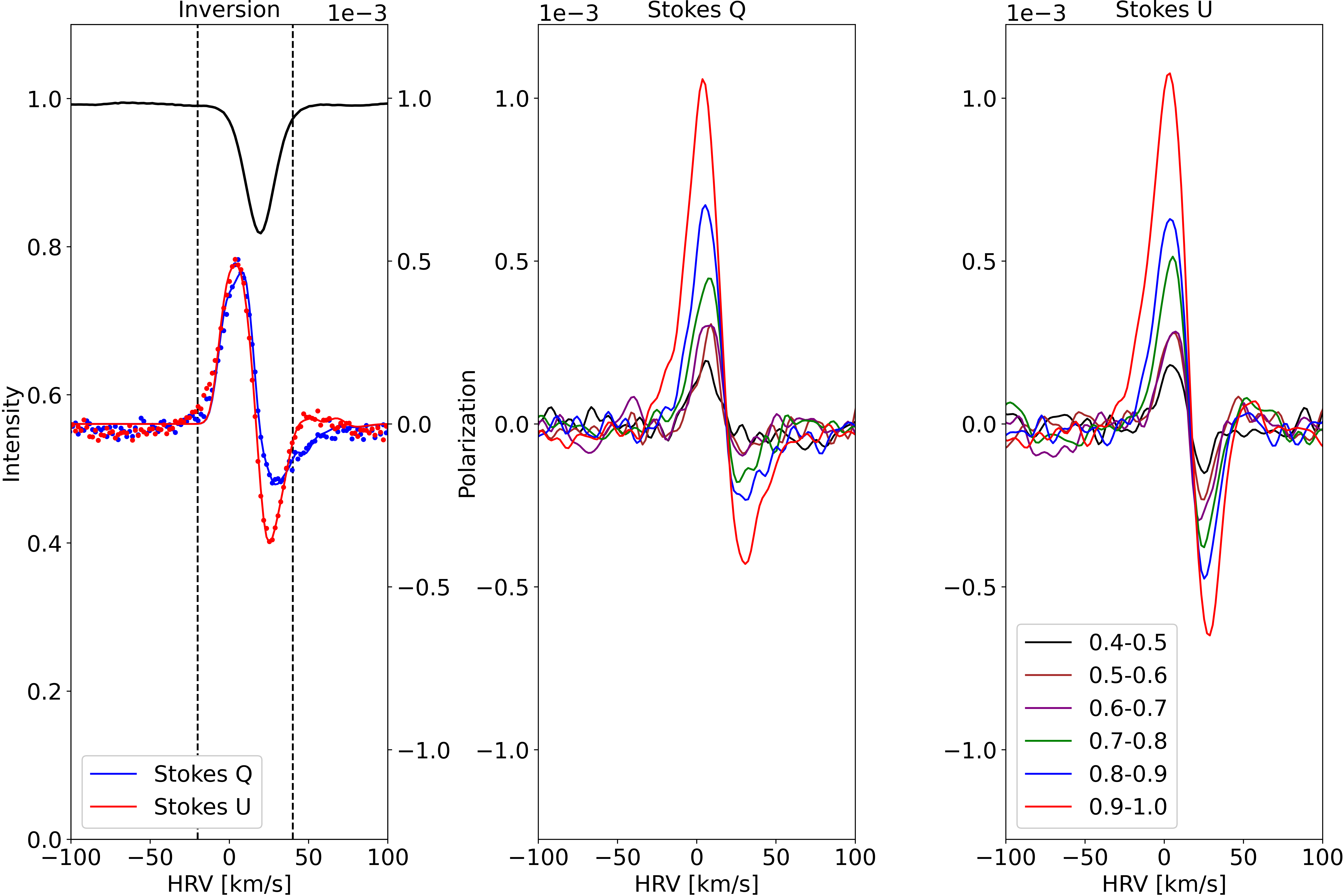}
\includegraphics[width=0.5\textwidth]{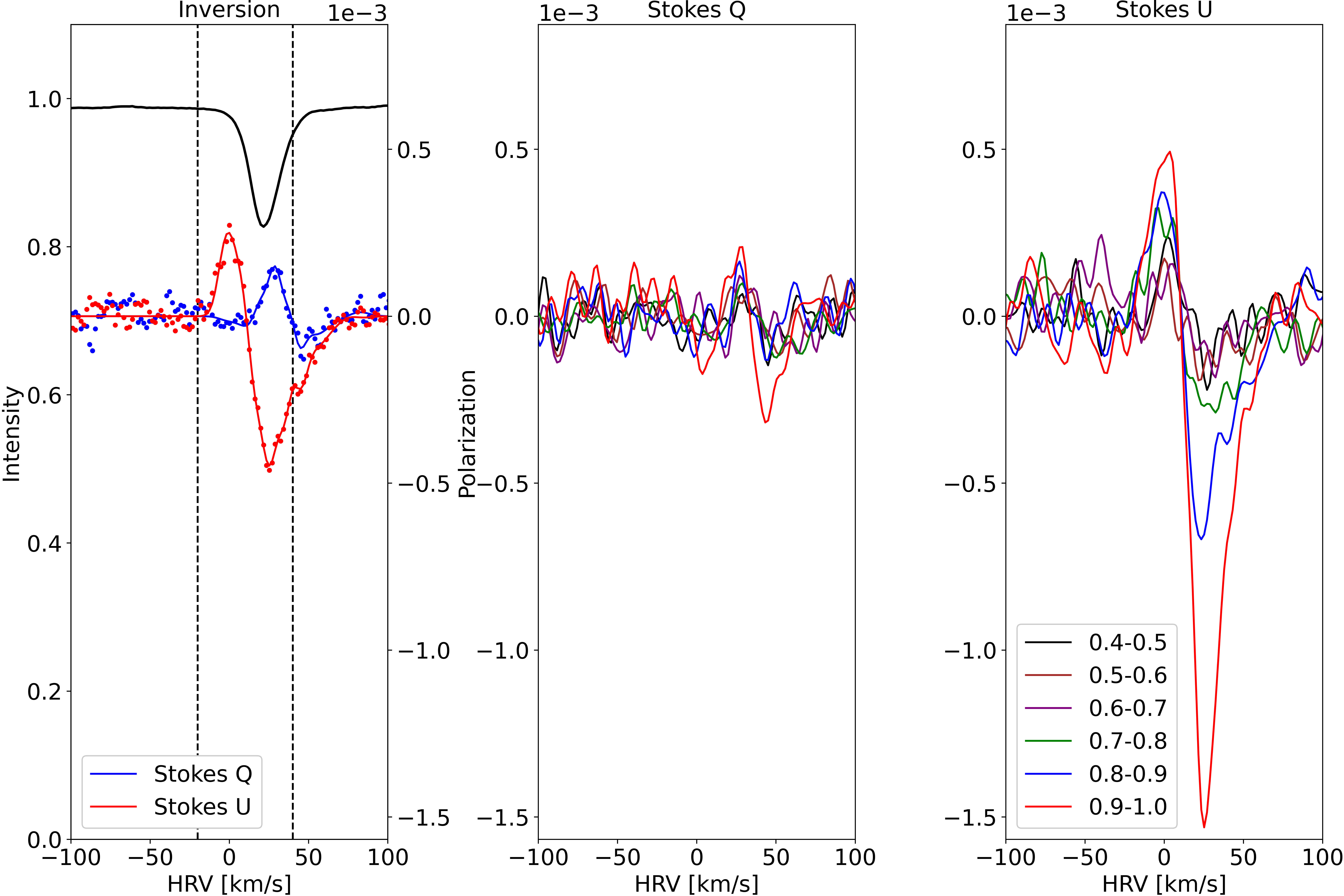}

\caption{   Stokes Q and U profiles for the two studied dates (December 12, 2013, above, and February  10, 2021, below). Stokes Q and U are defined in Narval and Neo-Narval such that positive Q corresponds to a linear polarization parallel to the direction of the celestial north at the position of the star.
Above and below, the left plot shows the observed Stokes Q and U profiles summed for the full spectral mask (blue and red dots respectively),  as well as the fit produced by the inversion algorithm (full lines). The black line shows the intensity line profile for reference. The dashed vertical lines indicate the adopted values of $V_*$ (redshifted one) and  $V_{max}$ (blueshifted one).
The center and right plots show the Stokes Q and U signals respectively, for each one of the masks grouping atomic lines of the spectra of Betelgeuse in terms of the depth of the line profile, from 0.4 to 1 in ranges of 0.1. }
\label{QUmasks}
\end{figure}

Figure\,\ref{QUmasks} shows the Stokes Q and U profiles for the dates of  December 20, 2013 (Narval instrument) and February 10, 2021 (Neo-Narval instrument).
 We performed LSD line addition  on the observed spectra using six different  masks that classify the atomic lines of the spectrum of Betelgeuse in terms of the depth of the intensity line. The six masks correspond to line depths  between 0.4 and 1  by steps of 0.1 (where the deepest lines have depths close to 1 and the shallowest ones close to 0.4). The details of these masks can be found in \cite{Auriere_2016}.
   As in Fig.\,4 of \cite{Auriere_2016}, in our Fig.\,\ref{QUmasks} the observed amplitudes of linear polarization at these dates appear to be well ordered, and the deeper the line the larger the signal. Over our whole dataset we see this tendency  broken in just a few cases: 14 out of 64 to be precise. But even when the signal is not observed to be strictly larger, it is seen to be equal, and never smaller, or otherwise hidden in noise, as in the $Q$ signal from February 2021.
 Noise also hides the smaller polarization signals of the shallower lines, with line depths smaller than 0.4, and because of this we  only consider  masks that stop at line depths of 0.4.
   
\begin{figure}[htbp]
\includegraphics[width=0.5\textwidth]{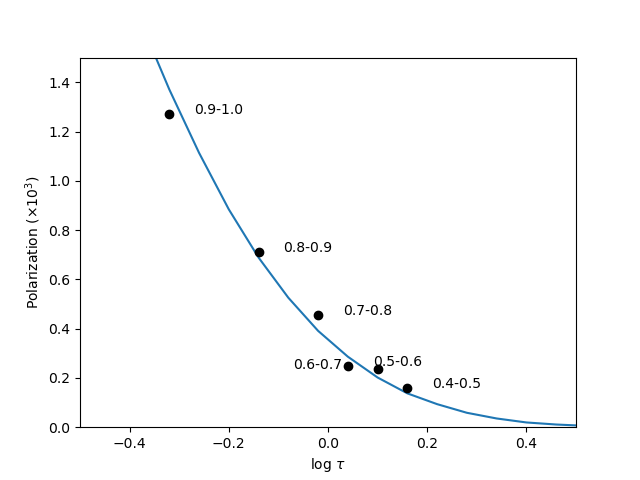}
\caption{Observed linear polarization  peak amplitudes (dots)  for the six masks on the polarization spectra, both Q and U, of December 12, 2013.  The abscissae of each dot is determined as the optical depth at which a gray atmosphere model fits the observed amplitude assuming a 1\% rate of local polarization.  The continuous line shows this model dependence.}
\label{Depth}
\end{figure}

To interpret this  order in the amplitude of the signal, we hypothesise that the observed polarization comes from just one scattering event in the formation of the continuum at that wavelength, followed by an immediate and in-place absorption of the polarized photon by an atom which re-emits it unpolarized. 
This unique event must happen at a given geometrical depth in the atmosphere of Betelgeuse. 
At this place, we can write a unique, simple form for the illumination moments in terms of the anisotropy factor $w_{\nu}$ and use it in Eq.\ref{grayeq2}. 
Assuming all other terms in  Eq.\ref{grayeq2} to be identical, we insist - as in Eq.\ref{fract} - in approximating the observed peak polarization amplitudes by the anisotropy factor $w_{\nu}$ and we plot this value at the optical depth at which it is predicted by a  gray atmosphere. Figure \ref{Depth}  shows the set of dots of  the amplitudes of observed linear polarization peaks at the optical depths thus assigned.
The gray atmosphere model, assuming this approximation of a single scattering event, allows us to easily interpret the observed ordering of the polarization signals in terms of height of formation:  the deeper the line, the higher its core forms in the atmosphere of Betelgeuse. It is tempting to look at the actual optical depths predicted by the grey atmosphere:  the deepest lines, with the largest linear polarization, would form at $\log \tau =-0.4$ while the shallowest lines in our analysis (line depths of 0.4 through 0.5) would form at 
$\log \tau=0.4$.   However, at the same time, we must be cautious about this crude determination of heights of formation. 

The ordering in height, and not the actual values, is the only information we use hereafter. We simply claim that certain lines form above certain other ones and that we can therefore build a 3D image made of layers, each layer corresponding to a mask filtered by the depth of the intensity line, following the order given in Fig.\,\ref{Depth}.   The two selected dates show the expected ordering of the signals and therefore allow us to order the masks in height and build a 3D image.  It must be said that not all the observations available and presented in 
 Table \ref{tab1} present the right ordering of the signals explained by our approximations. For those dates for which the ordering is not present, our model fails and we cannot assign a height to the masks, and so a  3D image is not  feasible. Nevertheless, these are not the object of the present work.

One may question whether or not one should  use the contribution functions of the intensity to improve our knowledge of the actual height of formation of different lines.
Our doubts about the interpretation of the intensity line profile raised above are a strong argument against using those contribution functions of the intensity. The imposition of the hypothesis of a single scattering event illuminated anisotropically as in a gray atmosphere keeps us from also using contribution functions for the polarization. The validity  of those approximations made so far is sufficient to order the heights of emergence of polarization, but not to go any further. We refrain from assigning any meaning to optical depths in the model, for example, or even trying to determine distances between layers.

 \subsection{Ambiguous solutions}
 \label{Section3}
 In spite of the large amount of information that is apparently available in the observed linear polarization spectra of Betelgeuse, it is not sufficient to produce a unique solution to the inverse problem of inferring an image which fits the observed spectrum. Part of this lack of unicity in the solution is due, as often, to the presence of noise. The broadening of the profiles, both instrumental and stellar (thermal, micro-, and macroturbulence broadenings would need to be considered), smears out the details and makes the algorithm particularly insensitive to the position and brightness of the dimmer granules. Also, linear polarization carries over its infamous 180-degree ambiguity \citep{LA18}, and  the whole inferred image can be rotated 180 degrees with no change  in the linear polarization profiles. One can also mirror the image with respect to any axis passing through the center of the stellar disk without changing the resulting polarization profiles.
 
When comparing Betelgeuse and CE\,Tau images to those inferred by interferometry, \cite{Auriere_2016} and \cite{LA18} were able to rely on the fact that bright structures at disk center are not changed by these ambiguities and that the brightest structures impose a preferred axis of symmetry.  This means that most of the images produced could be compared to interferometric analogs after a 180 degree rotation at most,  supporting the conclusion that both techniques were detecting the same photospheric structures.  
 
However, in general, multiple ambiguous solutions are available to the inversion algorithm, and if it is left to handle one observation independently of  those from other dates, there is no reason fto suggest that the solution found will preserve the same convention for the ambiguities of the other dates. We therefore select  one particular solution by hand at an initial date, usually one for which we have simultaneous interferometric images, and use this solution as the initial guess for the next date. We propagate one preferred solution among all the possible ones and are able to obtain a coherent picture from date to date.
 
 The same strategy is applied for the inversion of each one of the masks filtered by line depth at any date. The solutions for the mask with the deepest lines, which in the studied data has always the largest signals and therefore the largest S/N, is propagated as initial condition for the next date, but also as initial solution for the inversion of the next mask with shallower lines.
 
 This strategy produces coherent solutions from date to date, but  has a negative impact on the 3D images. As the solutions are propagated from line mask to line mask, from one geometrical depth to the next one, the same structures tend to appear in one layer upon the other. The algorithm will tend to produce column-like 3D structures. While these can be a welcome result in a convective scenario, we should remember that this is an implicit bias of the code.   On the positive side, we may claim that any structure that varies with geometrical depth in the inferred images can be trusted to be real:  the inversion algorithm is biased to reproduce the structures from the layer above, and so any change  is necessarily forced by the presence of information in the profiles, in defiance of that bias.
 
 \subsection{The relation between velocity and brightness}
Inferring a brightness distribution over the surface of Betelgeuse depends critically on retrieving spatial information on the polarized spectra. Simplistically, as described above, the ratio of Stokes Q to Stokes U gives information on the polar angle. The distance to disk center is recovered  by considering that all light-emitting plasma is moving vertically at the same speed. The projection of this speed onto the line of sight results in a linear relationship between distance to disk center and wavelength.  \cite{LA18} improved upon these rough rules by assuming a relationship between brightness and velocity, inspired by solar convection (see the appendix). All plasma moved vertically, but its velocity and sign depended on its brightness. The assumed relationship was that the brightest plasma, wherever it was over the disk, was rising at the maximum velocity $V_{max}$. This plasma would   contribute to the signal at  the wavelength given by the projection of that maximum velocity onto the line of sight. The darkest plasma, wherever it was, was sinking at this very same maximum velocity and  would contribute to the signal at the appropriate projected velocity.  All other plasma with intermediate brightness was given a velocity following a linear relationship such that the 25\% darkest plasma was sinking and the 75\% brightest was rising. The choice of this ratio of rising to sinking plasma was chosen by \cite{LA18} so that the resulting convective pattern showed a brightness contrast similar to the solar granulation. 

It is clear that this is a strong  constraint, but it is also questionable.
To start with, if sinking plasma occupies only a quarter of the surface, any mass conservation argument would lead to the proposal of sinking velocities that are larger than the rising ones, the ratio of both velocities being subject to density changes.
This is indeed what is observed in the Sun (see the Appendix and the references therein). 
But taking this fact into account would imply a new parameter to be retrieved by the inversion algorithm, the  sinking velocity,  without a significant amount of information left in the observations to constrain it. 
A better solution would be to infer this formal relationship  between velocity and brightness from numerical simulations, with the result that observations would be dependent on simulations and therefore unable to confirm them or rule them out. We are therefore left with this {\it ad hoc} relationship which, as in all the others above, must be considered before validating any conclusion.

\section{Three-dimensional imaging}

Making 3D images  simply requires application of  the inversion algorithm to each one of the LSD profiles resulting from masks filtered by the depth of the intensity line profile. 
The gray atmosphere and the single scattering approximation can ensure  that if the observed LSD profiles show  ordered amplitudes as described, with  deeper lines showing larger polarization amplitudes, we can order  the inferred images by height.
Although we cannot give an exact, or even approximate, optical depth for these layers, it appears that we are moving in the range $\log \tau = (0.4,-0.4)$. 
This is very near where the continuum forms. One of the basic assumptions as to the origin of the polarization signal, namely that we observe the depolarization of the continuum polarization, is thus independently verified. This allows us to say that, as long as the inferred images are valid, as confirmed by interferometry, they correspond to the photosphere of Betelgeuse at  those heights.

\begin{figure*}[htbp]
\includegraphics[width=0.3\textwidth]{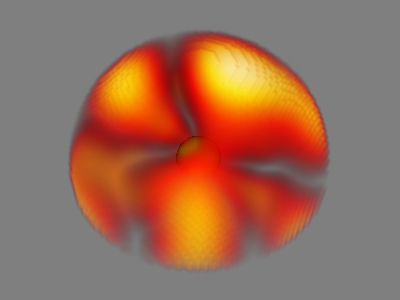}\includegraphics[width=0.3\textwidth]{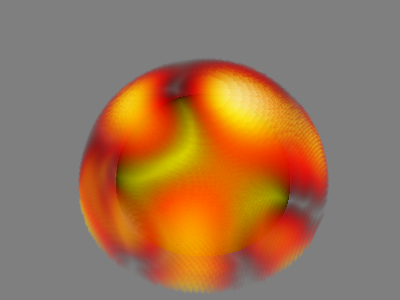}\includegraphics[width=0.3\textwidth]{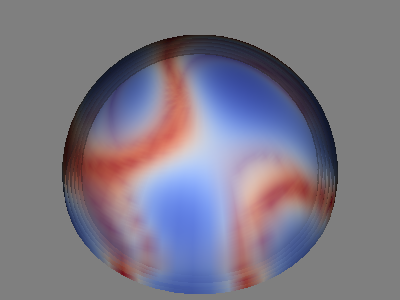}
\caption{Three-dimensional views of the photosphere of Betelgeuse on December, 20 2013. On the left, the explored region has been radially stretched to  ten times its probable size to better see the inferred vertical structures. The darkest regions have been made transparent. In the center, we show the star with the spatial scale that we consider closest to reality. In both images, a central yellow sphere marks the lower bound of our images, roughly corresponding to $\log \tau =1$, if this opacity were to be a smooth surface. On the right, radial velocities scaling from -60 to +60\,\kms; blue is rising plasma.}
\label{lms13}
\end{figure*}

\begin{figure*}[htbp]
\includegraphics[width=0.5\textwidth]{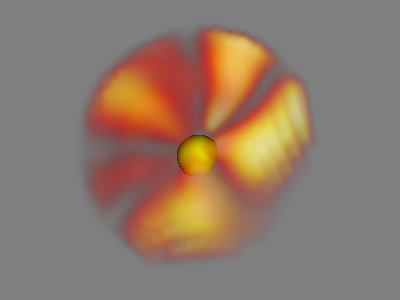}\includegraphics[width=0.5\textwidth]{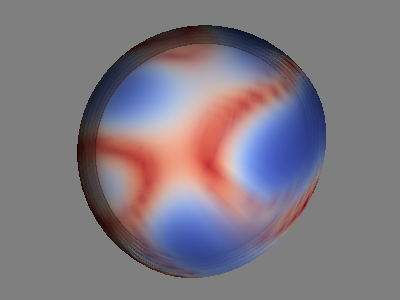}
\caption{Same a Fig.\ref{lms13} but for the night of February 10, 2021. Only the stretched version of the brightness image is shown.
We note that the ripples on some structures are due only to the graphical representation. }
\label{lms21}
\end{figure*}

For the present work, both selected dates from our observational data set show signal amplitudes with sufficiently large S/N, and that are well ordered when the signal is above noise. Our conditions to apply the listed approximations and build 3D images are therefore fulfilled. Figures \ref{lms13} and \ref{lms21} present the inferred 3D images for the 2013  Narval data and the 2021 Neo-Narval data, respectively.
In Fig.\ref{lms13} we present two  images of the brightness in three dimensions: one has a scale height related to what we believe best resembles  the right geometrical heights, while in the other the photosphere is stretched so that we can better see the inferred structures.   In Fig. \ref{lms21} we present only the stretched case.
We also produce, in both figures, a 3D image of the velocity along the radial direction which, we recall, is a monotonous function  of the brightness. 
Therefore, there is no information here that is not also present in the brightness images, except that sometimes looking at the data in different ways can reveal different aspects.  
Brightness in those images is just a measure of polarization amplitude. 
For any given layer, at  a constant height, the polarization amplitude is assumed constant and any change in the signal can only be attributed to changes in the emitted flux. 
Brightness in an image at constant height therefore represents true contrast of the emitted flux at that height. 
However, since the polarization amplitude increases with height,  each layer in the image is attributed to a correspondingly higher brightness than  the layer below, and no information is retrieved  as to the relationship between the emitted fluxes at different layers.

The upper limit  of the imaged photosphere is estimated to be between 1.1 and 1.3\,R$_*$, but this is just a rough  estimate. 
This height range is suggested, on its upper boundary, by the presence of the \textit{molsphere}\citep{perrin_interferometric_2004} at about 1.3\,R$_*$ which we assume marks the upper limit of formation of atomic lines in the mean atmosphere model of Betelgeuse. 
The lower boundary of 1.1\,R$_*$ is estimated from  ongoing studies on convective structures seen beyond  the limb of Betelgeuse and $\mu$\,Cep. The limb of the star acts as a geometrical horizon, limiting the minimum height of the observed structures \citep{LA22}. 
In optical depth, the gray atmosphere indicates that the linear polarization profiles are consistent with a region going from $\log \tau=0.4$ to -0.4. A solid sphere (yellow in the online version) in the center of the images indicates the surface $\log \tau =1$.

\section{Velocity profiles of the rising plasma}

As described in Sect. \ref{Section3}, in order to keep the coherence between the solutions of one layer and the contiguous ones, we are using, for the top layer, the solution for the top layer from the previous date; for the first date a random initial guess is used for this top layer.  
For any other layer, at a given date, the solution for the layer above is used as initial condition. 
This ensures that the layers show coherent structures, but biases the algorithm to produce  vertical structures that may be reminiscent of convective flows. 
Any such vertical structure in our images must be viewed with caution because of this bias in the inversion algorithm. 

Another approximation imposed in order to recover the 3D images is that the maximum vertical velocity of the plasma is exactly identical for all the layers. 
We describe this approximation  in some detail above. 
This maximum velocity is fixed once and for all after examination of the bluest wavelengths where signal can be observed (Fig.\,\ref{velsProf}). 
This gives us a global maximum velocity for all the observations in our data set, but not the actual maximum velocity of the plasma at any particular date. Furthermore, it is even more doubtful that, at any particular layer, the maximum velocity of the plasma is the same  as the layers above or below. Indeed, if the plasma is rising in a ballistic motion that is subject to gravity alone, one would expect that this maximum velocity decreases with height. As in most of the other approximations described above, we are forced to impose this one because we cannot afford to overwhelm the inversion algorithms with more free parameters. 

However, in the present case, this imposed approximation carries an unexpected benefit.   Once the maximum velocity of the plasma is fixed, the inversion code will interpret any polarization signal at the wavelength corresponding to this maximum velocity as light emitted at disk center. The brightest plasma over the disk will also be assigned such maximum vertical velocity but projected onto the line of sight if it happens that it is somewhere other than disk center, so that its projected velocity corresponds to the wavelength where the strongest signal is observed.  Now we can imagine a situation where, at the bottom layer, we have the brightest signal exactly at disk center. This plasma is moving vertically at  maximum speed, but its speed is decreasing with height. In the next layer above, this plasma, which is still at disk center, will be moving slower. Its polarization signal will be observed slightly  toward the red  with respect to the signal coming from the layer below. The inversion code, forced to conserve the same value of the maximum velocity for all layers, will infer  that this signal  is  not coming from  disk center anymore, but  from a slightly offset position. Layer upon layer, this situation is repeated: the plasma  in Betelgeuse will still be at disk center and moving vertically but at increasingly lower speeds; however, the inversion algorithm will place it farther and farther off disk center forced by an imposed constant maximum velocity. Consequently, the inferred 3D structure will bend toward the limb. The same argument will apply to  any other vertical flow anywhere over the disk as long as its velocity decreases with height.

This approximation of keeping the maximum velocity of the plasma independent of height will force any vertical flow subject to gravity to be inferred as a structure that bends toward the limb. From its curvature, we could infer the effective gravity force if only we were able to figure out its geometrical height. Conversely, we could derive the geometrical height of the structures if assured that gravity was the only force at work in this movement. 

If the plasma, rather than slowing down, was accelerating, the opposite argument would apply.  This cannot happen at disk center, because we have made sure when fixing the value of $V_{max}$ that no observed signal exceeds this value. However, it could happen for plasma off disk center whose projected velocity is below the threshold $V_{max}$.  In such a case, a true acceleration of the plasma in Betelgeuse would be seen as a polarization signal that, layer upon layer, shifts toward $V_{max}$. In that case, the inversion algorithm would infer vertical structures bending toward disk center. Finally, with this approximation at work, the only inferred  vertical structures would be those with a velocity that is constant with height.

\begin{figure*}[htbp]
\includegraphics[width=0.255\textwidth,angle=90]{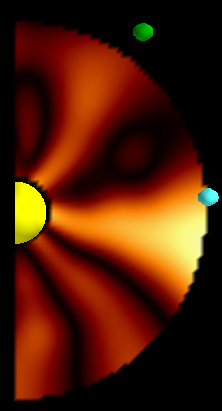}\includegraphics[width=0.5\textwidth]{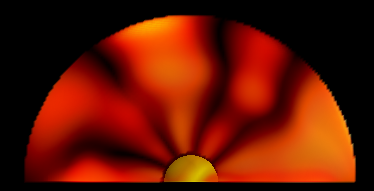}
\caption{2D cuts on the inferred photosphere of Betelgeuse on December 12,  2013 (left, Narval data)  and February 10, 2021 (right, NeoNarval data). Both cut planes are parallel to the central meridian plane of the disk. Several limb-bending structures are seen in one and the other images, which we interpret as plasma slowing down as it rises, subject to gravity. Also visible are two structures showing a radial profile, which we interpret as presenting a velocity  constant with height,  therefore implying the presence of a vertical force countering gravity. The colored dots identify the two   granules  out of the four followed in Fig. \ref{trajectos} which are visible in the image, though projected.}
\label{cuts}
\end{figure*}

Figure\,\ref{cuts} shows 2D cuts across the inferred images on the two selected dates.  In both cases, we can easily identify  structures bending toward the limb, and also structures rising radially. For the date of 2013, the image inferred from Narval data shows a disk center granule rising almost vertically if one looks at the brightest point at each layer (blue dot).  The two structures on the sides of this center granule are, on the other hand, clearly bending toward their respective limbs. One might see a coincidence in this blossoming of the structures as seen from Earth: the granules may, in reality, be bending and they are doing so symmetrically as seen from Earth. However, it appears simpler to accept that the inversion algorithm, forced to accept a maximum velocity constant with height is bending the structure in the inferred images in response to a decreasing velocity with height in Betelgeuse: these granules are simply following a ballistic motion, slowing down as they rise. However, the central granule shows a vertical structure. The symmetric expansion on both sides of the granule can be seen as a slowing down of some of its rising material. If all the material were slowing down, the center of the granule would dim with height, but in reality it  brightens. We must conclude that the inversion algorithm finds more and more polarization signal at the correct wavelength. There is always plasma moving at  maximum velocity at all heights, and so the velocity of the plasma is constant with height.

The data from Neo-Narval in 2021 provide further illustration of  a common observation over the 7 years of data, independent of the instrument used.  Once more, we see a limb-bending granule toward the left of the image, and a vertically rising granule near the center. This time the vertical granule is not exactly at disk center, yet both sides show a radial profile and the center of the granule is brighter with increasing height. We again conclude that the plasma in this structure is moving at constant speed independently of height.  The granule immediately to its right also merits attention; it shows a radial profile on its lower layers, and bends toward disk center at the top. Interpretation in terms of velocities would lead us to conclude that the plasma in the bottom of this structure is rising at a constant velocity for half of its height, and then accelerates in the top layers.

\begin{figure}[htbp]
\includegraphics[width=0.5\textwidth]{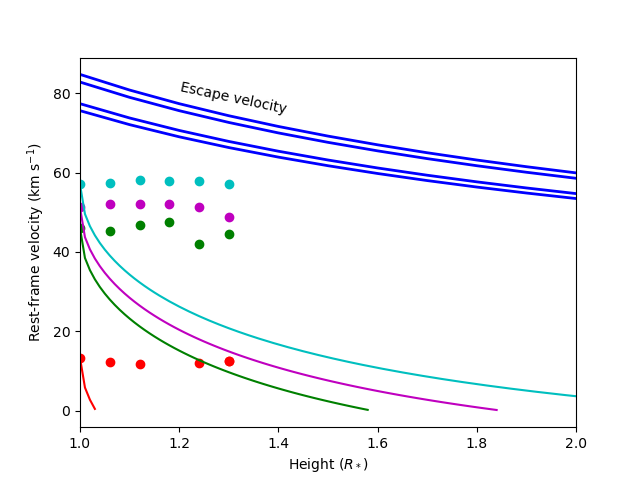}
\caption{Plasma velocities as a function of height in four granules from Dec 12, 2013. The dots show measured velocities at the center of four granules, two of which are identified in Fig.\,\ref{cuts}. These are assumed to be vertical structures and the velocity is computed by re-interpreting changes in the distance to disk center in terms of velocity. For comparison, the thin colored lines show the expected velocities in a ballistic scenario. The four thick blue lines show the escape velocity as a function of height for the four cases of mass 15 or 18\,M$_{\odot}$ and radius 955 or 1000\,R$_{\odot}$. }
\label{trajectos}
\end{figure}

\cite{LA18} inferred relatively high velocities of up to 40\,\kms \, for the rising plasma. 
Figure\,\ref{velsProf} confirms such high speeds and even allows for higher values at some particular dates, the two under study  in this work for example.  
Inspection of the 3D images suggests that this velocity is almost constant with height. 
Figure\,\ref{trajectos}  shows the velocities as a function of height for four granules in the image of December 12, 2013. 
To build this figure, we  retained  the  position of the brightest point at every layer inside several selected  granules, two of which are visible and marked in Fig.\,\ref{cuts}, and we propose  that this is a radial flow, that is, that in truth all those points are physically one above the other. 
Accepting this hypothesis implies that any change in the distance of the brightest point in the granule to the center of the disk when changing height must be attributed to a change in velocity and not a change in position. The measured position of the brightest point in the granule gives us a measure of the actual velocity at each layer. We computed these velocities as a function of  the layer that would leave the bright centers of each granule at the same position and plot them in Fig.\,\ref{trajectos}. 
We also plot  the escape velocities at each height, assuming masses of 15  and 18\,M$_{\odot}$ and radii of 955 and 1000\,R$_{\odot}$. 
Taking the initial velocity to be that  found in the deepest layer, we  also plot the expected velocities as a function of height if gravity were the only force acting on the plasma.  
The figure confirms the impression obtained from the cuts in the 3D image: the plasma is moving at roughly constant speeds with height. In retrospect, support to this conclusion may already be found in the positions of the peaks of polarization at different layers: with increasing height, the peaks appear at roughly the same wavelength, the same velocity or even an increasing velocity with height (Fig. \ref{QUmasks}). This observation is independent of all our approximations, but we were only able to reach the above conclusions once  the images inferred had allowed us to attribute those peaks in the spectral profiles to physical structures on the atmosphere of Betelgeuse. 

Two further conclusions are drawn from Fig.\,\ref{trajectos}. 
One is that  there must be at least one other force at work compensating  gravity almost perfectly to keep the velocity of the plasma constant with height. 
The buoyancy of the hot plasma makes it rise in any convective motion, and so part of this constant velocity must be due to this Archimedes' force. 
Nevertheless, numerical simulations indicate that this is not enough to keep velocities constant, and we must conclude that another force is at work.  

The second conclusion is that, on the top layers, this velocity may be very near the escape velocity for some of the observed plasma.
For those fast granules, it is sufficient that the force or forces counteracting gravity continue doing so for just a little longer for the  plasma to escape the gravity of Betelgeuse and efficiently contribute to the stellar wind. 
By selecting the highest value of our estimate of height span to draw Fig.\,\ref{trajectos} we show the most tantalizing scenario in which this plasma is quite near to  escaping gravity and contributing to the stellar wind. 
Our lower estimates of the maximum height at about $1.1\, R_*$, would still leave room for this plasma to slow down and return to the star, but this is without taking into consideration that, through examination of the time evolution of these granular structures in \cite{LA18}, we know that they are present for months at a time, and during that timescale the plasma does not seem to change its velocity significantly. As shown by \cite{Auriere_2016}, the polarization spectra from November 2013 through the end of January 2014 are quite similar in the velocity positions of the peaks as are the images inferred by \cite{LA18} for those dates. Three months at the measured velocities allow the plasma to travel well beyond 500$R_{\odot}$. 
If our measurements of the velocity of this plasma are valid through our long series of approximations, we captured it while escaping Betelgeuse and joining the stellar wind of this star.

\section{Conclusion}

Examination of the linear polarization in the atomic lines of the spectrum of Betelgeuse led \cite{LA18} to infer images of the photosphere of this red supergiant. In order to obtain  sufficiently high S/Ns, the polarization profiles of about 10 000 lines were added together.  However, we can also add polarization over smaller subsets of 1000 to 2000 atomic lines and still recover  signals. Here we classify them in groups according to their line depth, and obtain, for each date of our observation data set spanning 7 years, several mean polarization profiles as a function of line depth. \cite{Auriere_2016}  illustrated that the signals change from one group of lines to the other, indicating that these groups of lines contain different information on the physical conditions that lead to the emission of polarization.  

Here, we  examined, and in some cases justified, the many approximations and hypotheses that lead to the inference of images from these polarization signals. 
Two of them, the assumption of a gray atmosphere and the single scattering approximation, allow us to  also interpret the changing profiles seen in  lines with similar line depth. 
These two hypotheses together suggest that if the observed polarization amplitudes grow monotonically with line depth, one can  assign an optical height to each group of lines. 
The actual value of the optical depth may not be reliable, but the ordering is. 
This matches the rough expectation that the cores of deeper lines form higher in the atmosphere, 
though we reach this result without taking into account  the intensity line profiles themselves. 
Not all datasets collected  by Narval and Neo-Narval show the right ordering of polarization amplitudes that justify the two hypotheses of a gray atmosphere and single scattering. 
However, for the majority of  dates (50 out of 64), this condition applies and we can use the polarization profiles to infer a 3D image of the photosphere of Betelgeuse. 
Here we present 3D images of the photosphere of Betelgeuse for two particular dates in 2013 and 2021.

Given the many approximations and hypotheses required to infer such 3D images, any direct, filterless interpretation of them should be done with caution. The first date selected to create these 3D snapshots was also been studied by \cite{Auriere_2016} and \cite{LA18}. These authors showed that the inferred 2D image was comparable to interferometric images made close in time. This coincidence assigns a plausibility to the hypotheses used, and to  that  2D image.  The 3D images show how the observed granules form vertical coherent structures. This is somewhat expected, because, for a given layer, the inversion algorithm uses  the solution from the layer above as initial condition. Compelling as it is to interpret them as convection flows rising through the atmosphere, such an interpretation  must also be taken with caution. 

Nevertheless, the inferred images show features that are not implicit in the inversion algorithm and that, in our opinion, must be regarded as real. In the present work, we focus on the radial profile of those granules. The inversion algorithm uses as approximation an {\it a priori} maximum vertical velocity which is forced to be constant with height. Any plasma rising vertically through the photosphere and being slowed down by gravity will produce structures in the inferred image that bend toward the limb due to this approximation. We see several of them in the two investigated dates, but we also find examples of inferred vertical structures that do not bend. The best interpretation of this unexpected feature is that the plasma in these structures is rising at constant speed.

This result can be traced back to the observed polarization profiles: the points of maximum polarization  corresponding to those structures are seen to be at roughly the same wavelength independent of height. In our analysis, none of the involved hypotheses appear to be responsible for  this observation. Both the absence of an alternative explanation in terms of  biases in the inference algorithm and the presence of an observational feature supporting the result are tantalizing arguments in favor of the validity of our conclusion.

Plasma rising at constant velocity implies the presence of a force acting against gravity and compensating it. This force is already  present in the low photospheric heights covered by our data. 
We estimate that our top layer may be as high as 1.3\,R$_*$. At this maximum height, the observed velocities  are still below the escape limit, but only barely. If the force countering gravity acts for a little longer and allows the plasma to reach 1.6\,R$_*$ with this same velocity, the plasma will escape gravity and enter the realm of Betelgeuse stellar wind.    One or several forces must be counteracting gravity in the photosphere of Betelgeuse and maintaining plasma velocities up to large heights.   Escape velocities are within reach if only these forces act a little longer. The mechanisms for wind triggering in Betelgeuse and other red supergiants appear to be at work already in the photosphere.

\begin{acknowledgements}
S.G. acknowledges support under the Erasmus+ EU program for doctoral mobility. S.G. and R.K.-A. acknowledge partial support by the Bulgarian NSF project DN 18/2.
\end{acknowledgements}

\bibliographystyle{aa}

\bibliography{art71}

\begin{thebibliography}{33}
\expandafter\ifx\csname natexlab\endcsname\relax\def\natexlab#1{#1}\fi

\bibitem[{Auri{\`e}re {et~al.}(2010)Auri{\`e}re, Donati, Konstantinova-Antova,
  Perrin, Petit, \& Roudier}]{auriere_magnetic_2010}
Auri{\`e}re, M., Donati, J.-F., Konstantinova-Antova, R., {et~al.} 2010,
  Astronomy and Astrophysics, 516, L2

\bibitem[{Auri{\`e}re {et~al.}(2016)Auri{\`e}re, L{\'o}pez~Ariste, Mathias,
  L{\`e}bre, Josselin, Montarg{\`e}s, Petit, Chiavassa, Paletou, Fabas,
  Konstantinova-Antova, Donati, Grunhut, Wade, Herpin, Kervella, Perrin, \&
  Tessore}]{Auriere_2016}
Auri{\`e}re, M., L{\'o}pez~Ariste, A., Mathias, P., {et~al.} 2016, Astronomy
  and Astrophysics, 591, A119

\bibitem[{Bellot~Rubio(2009)}]{2009ApJ...700..284B}
Bellot~Rubio, L.~R. 2009, The Astrophysical Journal, 700, 284, aDS Bibcode:
  2009ApJ...700..284B

\bibitem[{Bertout \& Magnan(1987)}]{bertout_line_1987}
Bertout, C. \& Magnan, C. 1987, Astronomy and Astrophysics, 183, 319

\bibitem[{Chandrasekhar(1945)}]{chandrasekhar_formation_1945}
Chandrasekhar, S. 1945, Reviews of Modern Physics, 17, 138

\bibitem[{Chiavassa {et~al.}(2011)Chiavassa, Freytag, Masseron, \&
  Plez}]{chiavassa_radiative_2011}
Chiavassa, A., Freytag, B., Masseron, T., \& Plez, B. 2011, Astronomy and
  Astrophysics, 535, A22

\bibitem[{Doherty(1986{\natexlab{a}})}]{1986ApJ...307..261D}
Doherty, L.~R. 1986{\natexlab{a}}, The Astrophysical Journal, 307, 261, aDS
  Bibcode: 1986ApJ...307..261D

\bibitem[{Doherty(1986{\natexlab{b}})}]{doherty_polarization_1986}
Doherty, L.~R. 1986{\natexlab{b}}, The Astrophysical Journal, 307, 261

\bibitem[{Donati {et~al.}(2006)Donati, Catala, Landstreet, \&
  Petit}]{Donati2006}
Donati, J.~F., Catala, C., Landstreet, J.~D., \& Petit, P. 2006, 358, 362,
  conference Name: Solar Polarization 4 ADS Bibcode: 2006ASPC..358..362D

\bibitem[{Donati {et~al.}(1997)Donati, Semel, Carter, Rees, \&
  Collier~Cameron}]{donati_spectropolarimetric_1997}
Donati, J.-F., Semel, M., Carter, B.~D., Rees, D.~E., \& Collier~Cameron, A.
  1997, Monthly Notices of the Royal Astronomical Society, 291, 658

\bibitem[{Freytag {et~al.}(2002)Freytag, Steffen, \&
  Dorch}]{freytag_spots_2002}
Freytag, B., Steffen, M., \& Dorch, B. 2002, Astronomische Nachrichten, 323,
  213

\bibitem[{Gray(2008)}]{Gray2008}
Gray, D.~F. 2008, The Astronomical Journal, 135, 1450, aDS Bibcode:
  2008AJ....135.1450G

\bibitem[{Haubois {et~al.}(2009)Haubois, Perrin, Lacour, Verhoelst, Meimon,
  Mugnier, Thi{\'{e}}baut, Berger, Ridgway, Monnier, Millan-Gabet, \&
  Traub}]{Haubois2009}
Haubois, X., Perrin, G., Lacour, S., {et~al.} 2009, Astronomy and Astrophysics,
  508, 923

\bibitem[{Josselin {et~al.}(2015)Josselin, Lambert, Auri{\`e}re, Petit, \&
  Ryde}]{josselin_atmospheres_2015}
Josselin, E., Lambert, J., Auri{\`e}re, M., Petit, P., \& Ryde, N. 2015,
  Proceedings of the International Astronomical Union , Volume 10 , Symposium
  S305: Polarimetry: From the Sun to Stars and Stellar Environments , December
  2014, 305, 299, conference Name: Polarimetry

\bibitem[{Kervella {et~al.}(2018)Kervella, Decin, Richards, Harper, McDonald,
  O'Gorman, Montarg{\`e}s, Homan, \& Ohnaka}]{kervella_close_2018}
Kervella, P., Decin, L., Richards, A. M.~S., {et~al.} 2018, Astronomy and
  Astrophysics, 609, A67

\bibitem[{Koza {et~al.}(2006)Koza, Ku{\v c}era, Ryb{\'a}k, \&
  W{\"o}hl}]{koza_photospheric_2006}
Koza, J., Ku{\v c}era, A., Ryb{\'a}k, J., \& W{\"o}hl, H. 2006, Astronomy and
  Astrophysics, Volume 458, Issue 3, November II 2006, pp.941-951, 458, 941

\bibitem[{Kravchenko {et~al.}(2018)Kravchenko, Van~Eck, Chiavassa, Jorissen,
  Freytag, \& Plez}]{kravchenko_tomography_2018}
Kravchenko, K., Van~Eck, S., Chiavassa, A., {et~al.} 2018, Astronomy and
  Astrophysics, 610, A29

\bibitem[{Landi~Degl'Innocenti \&
  Landolfi(2004)}]{landi_deglinnocenti_polarization_2004}
Landi~Degl'Innocenti, E. \& Landolfi, M. 2004, Polarization in {Spectral}
  {Lines}, Vol. 307 (Kluwer Academic Publishers)

\bibitem[{L{\'o}pez~Ariste {et~al.}(2018)L{\'o}pez~Ariste, Mathias, Tessore,
  L{\`e}bre, Auri{\`e}re, Petit, Ikhenache, Josselin, Morin, \&
  Montarg{\`e}s}]{LA18}
L{\'o}pez~Ariste, A., Mathias, P., Tessore, B., {et~al.} 2018, Astronomy and
  Astrophysics, 620, A199

\bibitem[{L{\'o}pez~Ariste {et~al.}(2019)L{\'o}pez~Ariste, Tessore,
  Carl{\'\i}n, Mathias, L{\`e}bre, Morin, Petit, Auri{\`e}re, Gillet, \&
  Herpin}]{lopez_ariste_asymmetric_2019}
L{\'o}pez~Ariste, A., Tessore, B., Carl{\'\i}n, E.~S., {et~al.} 2019, Astronomy
  and Astrophysics, 632, A30

\bibitem[{L{\'o}pez~Ariste(2022)}]{LA22}
L{\'o}pez~Ariste, A. e.~a. 2022, In Preparation

\bibitem[{Malherbe {et~al.}(2012)Malherbe, Roudier, Rieutord, Berger, \&
  Franck}]{malherbe_acoustic_2012}
Malherbe, J.-M., Roudier, T., Rieutord, M., Berger, T., \& Franck, Z. 2012,
  Solar Physics, 278, 241

\bibitem[{Mathias {et~al.}(2018)Mathias, Auri{\`e}re, L{\'o}pez~Ariste, Petit,
  Tessore, Josselin, L{\`e}bre, Morin, Wade, Herpin, Chiavassa, Montarg{\`e}s,
  Konstantinova-Antova, Kervella, Perrin, Donati, \& Grunhut}]{Mathias:2018aa}
Mathias, P., Auri{\`e}re, M., L{\'o}pez~Ariste, A., {et~al.} 2018, Astronomy
  and Astrophysics, 615, A116

\bibitem[{Montarg{\`e}s {et~al.}(2021)Montarg{\`e}s, Cannon, Lagadec, de~Koter,
  Kervella, Sanchez-Bermudez, Paladini, Cantalloube, Decin, Scicluna,
  Kravchenko, Dupree, Ridgway, Wittkowski, Anugu, Norris, Rau, Perrin,
  Chiavassa, Kraus, Monnier, Millour, Le~Bouquin, Haubois, Lopez, Stee, \&
  Danchi}]{montarges_dimming_2021}
Montarg{\`e}s, M., Cannon, E., Lagadec, E., {et~al.} 2021, Nature, 594, 365,
  aDS Bibcode: 2021Natur.594..365M

\bibitem[{Montarg{\`e}s {et~al.}(2016)Montarg{\`e}s, Kervella, Perrin,
  Chiavassa, Bouquin, B, Auri{\`e}re, L{\'o}pez~Ariste, Mathias, Ridgway,
  Lacour, Haubois, \& Berger}]{montarges_close_2016}
Montarg{\`e}s, M., Kervella, P., Perrin, G., {et~al.} 2016, Astronomy and
  Astrophysics, 588, A130

\bibitem[{Montarg{\`e}s {et~al.}(2018)Montarg{\`e}s, Norris, Chiavassa,
  Tessore, L{\`e}bre, \& Baron}]{montarges_convective_2018}
Montarg{\`e}s, M., Norris, R., Chiavassa, A., {et~al.} 2018, Astronomy and
  Astrophysics, 614, A12

\bibitem[{Nesis {et~al.}(1992)Nesis, Bogdan, Cattaneo, Hanslmeier, Knoelker, \&
  Malagoli}]{1992ApJ...399L..99N}
Nesis, A., Bogdan, T.~J., Cattaneo, F., {et~al.} 1992, The Astrophysical
  Journal, 399, L99, aDS Bibcode: 1992ApJ...399L..99N

\bibitem[{Oba {et~al.}(2017)Oba, Iida, \& Shimizu}]{oba_height-dependent_2017}
Oba, T., Iida, Y., \& Shimizu, T. 2017, The Astrophysical Journal, 836, 40

\bibitem[{Perrin {et~al.}(2004)Perrin, Ridgway, Coud{\'e}~du Foresto,
  Mennesson, Traub, \& Lacasse}]{perrin_interferometric_2004}
Perrin, G., Ridgway, S.~T., Coud{\'e}~du Foresto, V., {et~al.} 2004, Astronomy
  and Astrophysics, 418, 675

\bibitem[{Schwarz \& Clarke(1984)}]{schwarz_polarization_1984}
Schwarz, H.~E. \& Clarke, D. 1984, Astronomy and Astrophysics, Vol. 132, p.
  370-374 (1984), 132, 370

\bibitem[{Solanki {et~al.}(1996)Solanki, Rueedi, Bianda, \&
  Steffen}]{solanki_detection_1996}
Solanki, S.~K., Rueedi, I., Bianda, M., \& Steffen, M. 1996, Astronomy and
  Astrophysics, 308, 623

\bibitem[{Stein \& Nordlund(1998)}]{1998ApJ...499..914S}
Stein, R.~F. \& Nordlund, {\AA}. 1998, The Astrophysical Journal, 499, 914, aDS
  Bibcode: 1998ApJ...499..914S

\bibitem[{Uitenbroek {et~al.}(1998)Uitenbroek, Dupree, \&
  Gilliland}]{uitenbroek_spatially_1998}
Uitenbroek, H., Dupree, A.~K., \& Gilliland, R.~L. 1998, The Astronomical
  Journal, 116, 2501

\end{thebibliography}

 \begin{appendix}
 
 \section{Brightness versus velocity: the solar case}
 Both \cite{LA18} and the present work include a relationship between the velocity of the plasma and its brightness in the model used to fit the observed linear polarization profiles and infer the images (2D or 3D) of the photosphere of Betelgeuse.  It is obvious that such a relationship must exist in a convective atmosphere: the hot and bright plasma rises in the atmosphere; as it cools down it slows down and is advected horizontally toward the edges of the convective cell; the cold and dark plasma sinks toward the inner layers of the atmosphere.  
 
 \cite{koza_photospheric_2006} and \cite{oba_height-dependent_2017}  published such  relationships for  the solar granulation. The typical values of the solar convection appear in that data. The rising velocity of the hot plasma can be as high as 2 \kms. Similar values are seen for the sinking velocity, although supersonic velocities have often been measured in the intergranular lanes \citep{2009ApJ...700..284B,1992ApJ...399L..99N,solanki_detection_1996} and in general these sinking velocities are larger than the velocities of the rising plasma. Indeed, \cite{koza_photospheric_2006} compare their measurements with numerical simulations by \cite{1998ApJ...499..914S}, which show  higher sinking velocities apparently below the formation region of the spectral line used to measure the solar granulation.
 
\begin{figure}[htbp]
\includegraphics[width=0.5\textwidth]{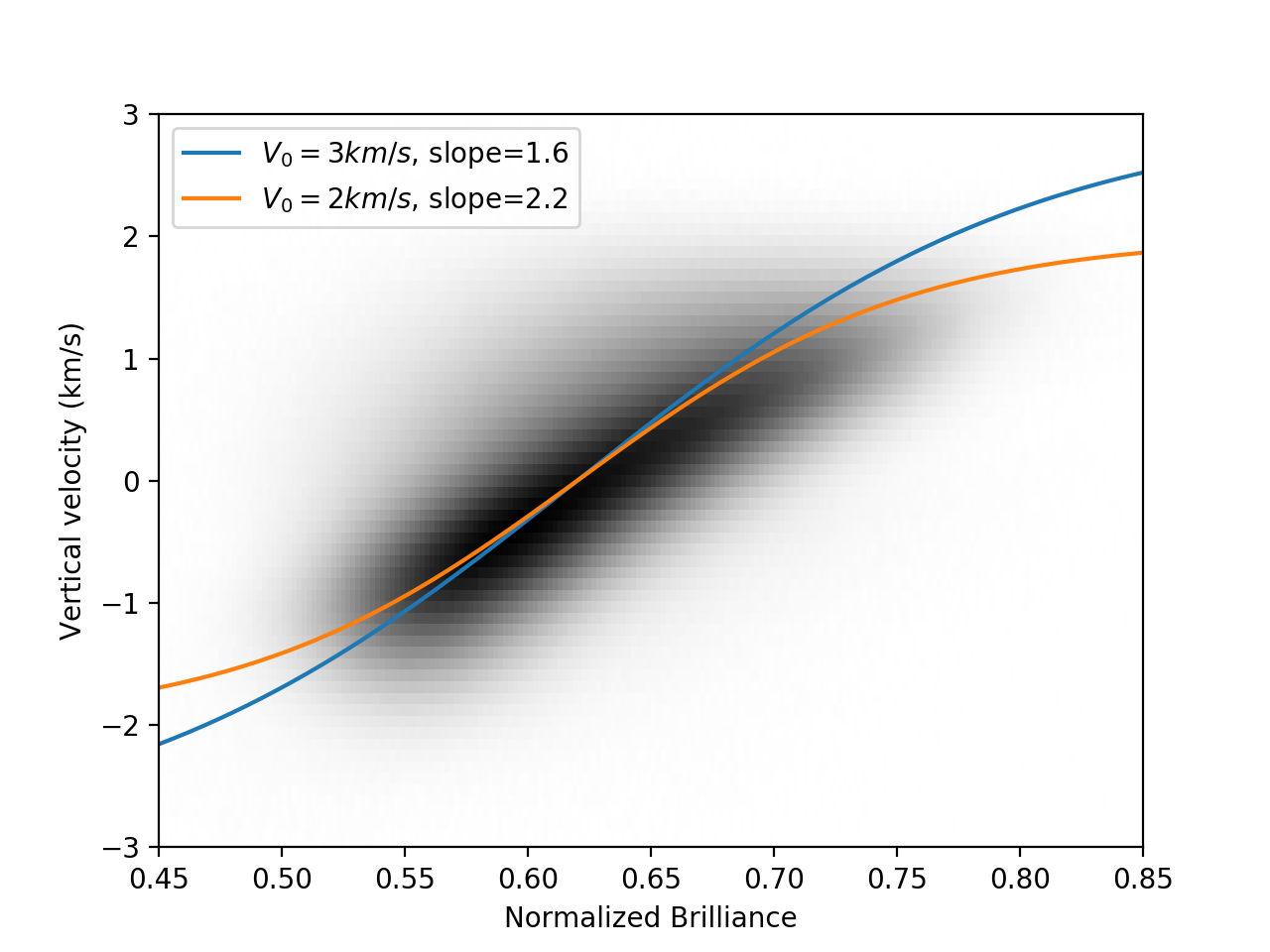}
\caption{ Histogram of pairs of values of radial velocity and relative intensity in the solar granulation using a dataset observed by Hinode \citep{malherbe_acoustic_2012}. Two hyperbolic tangents are plotted over the histogram to illustrate how this function can capture the relationship between the two observables. }
\label{Sol}
\end{figure}

 In order to test the functional form given to this relationship for Betelgeuse, we made our own measurements using a dataset of solar granulation observed with Hinode and treated and described by \cite{malherbe_acoustic_2012}.  The time series of the patch of solar granulation contain measurements of both the emergent intensity in the continuum near the \ion{Fe}{I} line at 557.6nm and the velocity along the line of sight. Because of the position of the observed solar region near disk center, this velocity can be safely seen as the radial velocity of the plasma. Figure \ref{Sol} shows a 2D histogram of the number of points for each pair of values of intensity and velocity.
 
 Plotted over the histogram in FIg.\ref{Sol} appear two curves following  the function:
 \begin{equation}
 V=V_0 \tanh (slope (B-0.62)/0.3)
 \end{equation}
 where the values of the parameters $V_0$ and $slope$ are given in the figure.  $V$ and $B$ are the radial velocity and the intensity of the solar granulation respectively. The two other numerical parameters simply make sure that the curve is well centered and suits  the relative scale used for brightness. The use of the hyperbolic tangent is justified as a function which is linear in its core but which saturates at large values. This behavior describes the quasi-linear relationship between brightness and velocity, while avoiding numerical issues if  anomalously large values were to appear in the data.   
 
 The nice fit of this hyperbolic tangent functional relationship to the solar granulation is the sole justification of the use of the same function for Betelgeuse, adapting the numerical parameters to the velocities of up to 60 \kms\, observed in its photosphere.
 \onecolumn
 \section{Log of Observations}

 \begin{longtable}{lccccc}
 \caption{Log of Narval  and Neo-Narval  observations of Betelgeuse and polarimetric measurements since August 2018.}\\
\hline 
\hline
Date &  Julian date & Stokes & Number of cycles  &  Narval (N) &  Peak SNR \\ 
& & & averaged & NeoNarval (NN) &  \\
\hline
\endfirsthead
\caption{continued}\\
\hline
\hline
Date & Julian date & Stokes & Number of cycles & Narval (N) &  Peak SNR \\
 & & & averaged & NeoNarval (NN) &  \\
\hline
\endhead
\multicolumn{6}{r}{\textit{Continued on next page}} \\
\endfoot
\endlastfoot
August    18, 2018 & 8349.648 & Q & 2 &  N  & 837\\
August    18, 2018 & 8349.654 & U & 2 &  N  & 1223\\
September 19, 2018 & 8381.675 & Q & 2 &  N  & 1418\\
September 19, 2018 & 8381.68 & U & 2 &  N  & 1330\\
October   05, 2018 & 8397.694 & Q & 2 &  N  & 1509\\
October   05, 2018 & 8397.7 & U & 2 &  N  & 1562\\
October   23, 2018 & 8415.635 & Q & 2 &  N  & 1280\\
October   23, 2018 & 8415.64 & U & 2 &  N  & 1293\\
November  13, 2018 & 8436.59 & Q & 2 &  N  & 1075\\
November  13, 2018 & 8436.596 & U & 2 &  N  & 992\\
January   04, 2019 & 8488.53 & Q & 2 &  N  & 1180\\
January   04, 2019 & 8488.535 & U & 2 &  N  & 1112\\
January   15, 2019 & 8499.432 & Q & 2 &  N  & 1107\\
January   15, 2019 & 8499.436 & U & 2 &  N  & 1163\\
January   21, 2019 & 8505.427 & Q & 2 &  N  & 473\\
January   21, 2019 & 8505.433 & U & 2 &  N  & 492\\
January   26, 2019 & 8510.458 & Q & 2 &  N  & 1224\\
January   26, 2019 & 8510.463 & U & 2 &  N  & 1279\\
March     11, 2019 & 8554.379 & Q & 2 &  N  & 1441\\
March     11, 2019 & 8554.386 & U & 2 &  N  & 1512\\
March     28, 2019 & 8571.365 & Q & 2 &  N  & 1583\\
March     28, 2019 & 8571.37 & U & 2 &  N  & 1661\\
April     27, 2019 & 8601.357 & Q & 2 &  N  & 1680\\
April     27, 2019 & 8601.366 & U & 2 &  N  & 1644\\
April     30, 2019 & 8604.332 & Q & 2 &  N  & 397\\
April     30, 2019 & 8604.341 & U & 2 &  N  & 974\\
August    20, 2019 & 8716.671 & Q & 2 &  N  & 966\\
August    22, 2019 & 8718.676 & U & 2 &  N  & 1341\\
February  02, 2020 & 8882.332 & Q & 2 & NN  & 688\\
February  02, 2020 & 8882.33 & U & 2 & NN  & 805\\
February  14, 2020 & 8894.369 & Q & 2 & NN  & 805\\
February  14, 2020 & 8894.366 & U & 2 & NN  & 855\\
February  21, 2020 & 8901.378 & Q & 2 & NN  & 929\\
February  21, 2020 & 8901.381 & U & 2 & NN  & 880\\
February  24, 2020 & 8904.381 & Q & 2 & NN  & 957\\
February  24, 2020 & 8904.384 & U & 2 & NN  & 1028\\
March     11, 2020 & 8920.289 & Q & 2 & NN  & 847\\
March     11, 2020 & 8920.287 & U & 2 & NN  & 947\\
August    22, 2020 & 9084.665 & Q & 2 & NN  & 1152\\
August    22, 2020 & 9084.662 & U & 2 & NN  & 958\\
September 03, 2020 & 9096.654 & Q & 2 & NN  & 652\\
September 03, 2020 & 9096.658 & U & 2 & NN  & 611\\
September 29, 2020 & 9122.699 & Q & 2 & NN  & 882\\
September 29, 2020 & 9122.701 & U & 2 & NN  & 831\\
October   17, 2020 & 9140.667 & Q & 2 & NN  & 839\\
October   17, 2020 & 9140.668 & U & 2 & NN  & 936\\
October   30, 2020 & 9153.605 & Q & 2 & NN  & 903\\
October   30, 2020 & 9153.607 & U & 2 & NN  & 897\\
November  23, 2020 & 9177.468 & Q & 2 & NN  & 550\\
November  23, 2020 & 9177.469 & U & 2 & NN  & 541\\
December  17, 2020 & 9201.413 & Q & 2 & NN  & 829\\
\pagebreak
December  17, 2020 & 9201.415 & U & 2 & NN  & 813\\
January   06, 2021 & 9221.4 & Q & 2 & NN  & 996\\
January   06, 2021 & 9221.403 & U & 2 & NN  & 935\\
February  10, 2021 & 9256.418 & Q & 2 & NN  & 1035\\
February  10, 2021 & 9256.421 & U & 2 & NN  & 977\\[.5cm]
\label{tab1}
\end{longtable}
\textbf{Notes:} Columns give the date, the heliocentric Julian date (+2\,450\,000), the observed Stokes parameter, the number of Stokes $QU$ collected, and the  instrument used (N stands for Narval, NN for Neo-Narval). An observation consists upon 4 exposures with changing polarimetric modulation that, after reduction, produce polarization spectra of either Stokes $Q$ or $U$. Previous  $QU$ measurements are summarised by \cite{Auriere_2016,Mathias:2018aa} and \cite{LA18}. Beyond a 2-year proprietary embargo, all data is publicly available at PolarBase (http://polarbase.irap.omp.eu/).

 \end{appendix}

\end{document}